%% file: main.tex
\documentclass[sigconf]{acmart}

\usepackage{hyperref}
\usepackage{multirow}
\usepackage[utf8]{inputenc}
\usepackage[ruled, lined, longend, linesnumbered]{algorithm2e}
\usepackage{amsmath,multicol}
\usepackage{amssymb}

\usepackage{array}
\newcolumntype{L}[1]{>{\raggedright\let\newline\\\arraybackslash\hspace{0pt}}m{#1}}
\newcolumntype{C}[1]{>{\centering\let\newline\\\arraybackslash\hspace{0pt}}m{#1}}
\newcolumntype{R}[1]{>{\raggedleft\let\newline\\\arraybackslash\hspace{0pt}}m{#1}}

\input{header}

\newcommand{\revise}[1]{{{#1}}}

\copyrightyear{2020} 
\acmYear{2020} 
\setcopyright{acmcopyright}\acmConference[MobiSys '20]{The 18th Annual International Conference on Mobile Systems, Applications, and Services}{June 15--19, 2020}{Toronto, ON, Canada}
\acmBooktitle{The 18th Annual International Conference on Mobile Systems, Applications, and Services (MobiSys '20), June 15--19, 2020, Toronto, ON, Canada}
\acmPrice{15.00}
\acmDOI{10.1145/3386901.3388948}
\acmISBN{978-1-4503-7954-0/20/06}

\begin{document}
	
	\title{Approximate Query Service on Autonomous IoT Cameras}
	
	\author{Mengwei Xu}
	\authornote{Work performed while visiting Purdue}
	\affiliation{%
		\institution{Peking University}
	}
	\email{mwx@pku.edu.cn}
	
	\author{Xiwen Zhang}
	\affiliation{%
		\institution{Purdue ECE}
	}
	\email{zhan2977@purdue.edu}
	
	\author{Yunxin Liu}
	\affiliation{%
		\institution{Microsoft Research}
	}
	\email{yunxin.liu@microsoft.com}

	\author{Gang Huang}
	\affiliation{%
		\institution{Peking University}
	}
	\email{hg@pku.edu.cn}
	
	\author{Xuanzhe Liu}
	\authornote{Corresponding author}
	\affiliation{%
		\institution{Peking University}
	}
	\email{xzl@pku.edu.cn}
	
	\author{Felix Xiaozhu Lin}
	\affiliation{%
		\institution{Purdue ECE}
	}
	\email{xzl@purdue.edu}
	
	\input{abstract}


	
	\maketitle

\input{todo}\input{intro}
	\input{bkng}

\input{overview}
\input{design}
	\input{impl}

\input{eval}

	\input{related}
	\input{conclusion}

	\section*{Acknowledgment}
	We thank the anonymous reviewers and our shepherd, Dr. Aakanksha Chowdhery, for their valuable feedback.
	The authors affiliated with Peking University were
	supported by the National Key R\&D Program of China under the grant number 2018YFB1004800,
	the National Natural Science Foundation of China under grant number 61725201,
	the R\&D projects in key areas of Guangdong Province under grant number 2020B010164002,
	the Beijing Outstanding Young Scientist Program under grant number BJJWZYJH01201910001004,
	and partially supported by the Key Laboratory of Intelligent Passenger Service of Civil Aviation.
	The authors thank NVIDIA for GPU donation.
	\bibliographystyle{ACM-Reference-Format}
	\bibliography{bib/abr-short,bib/xzl,bib/cv,bib/book,mwx}
	
\end{document}

%% file: header.tex
\usepackage{wrapfig}
\usepackage{comment}
\usepackage{colortbl}
\usepackage{tabularx}

\usepackage{color,soul}
\usepackage{graphics}
\usepackage{graphicx}
\usepackage{lipsum}

\usepackage{CJKutf8} 

\usepackage{tikz}
\usetikzlibrary{decorations.pathmorphing, patterns,shapes,snakes} 

\usepackage[inline]{enumitem}	
\usepackage{listings} 
\usepackage{lipsum}

\usepackage{capt-of}	

\usepackage{ulem}  

\ifdefined\HCode
\usepackage[compatibility=false]{caption}

\else
\usepackage{subcaption} 
\fi

\usepackage{mathrsfs}	
\usepackage{amsfonts}

\usepackage[export]{adjustbox} 

\definecolor{myForestGreen}{RGB}{34, 139, 34}
\definecolor{airforceblue}{rgb}{0.36, 0.54, 0.66}
\newcommand*\circled[1]{\tikz[baseline=(char.base)]{
		\node[shape=circle,fill=black,draw,text=white,inner sep=1pt] (char) {#1};}}

\renewcommand{\paragraph}[1]{\vskip 3pt\noindent\textbf{#1 }}	 

  {\begin{list}{$\bullet$}%
     {\setlength{\parsep}{0pt}%
      \setlength{\topsep}{0pt}%
      \setlength{\itemsep}{2pt}}}%
  {\end{list}}
\newcommand\note[1]{\sethlcolor{yellow} \hl{#1}} 
\newcommand\Note[1]{\sethlcolor{green} \hl{#1}} 

\newcommand{\textbfit}[1]{\textbf{\textit{#1}}}
		
\newcommand\sect[1]{Section~\ref{sec:#1}}	
		
\newenvironment{myitemize}%
  {\begin{itemize}
	[leftmargin=0cm,
		itemindent=.3cm,
		labelwidth=\itemindent,
		labelsep=0pt,
		parsep=3pt,
		topsep=2pt,
		itemsep=1pt,
		align=left]
  }%
  {\end{itemize}}    

  {\begin{enumerate}
	[leftmargin=0cm,itemindent=.5cm,labelwidth=\itemindent,
		labelsep=0pt,
		parsep=1pt,
		topsep=1pt,
		itemsep=3pt,
		align=left]
  }%
  {\end{enumerate}}    
  
  {\begin{enumerate*}
	[label=\roman*)]
  }%
  {\end{enumerate*}}      

  {\begin{enumerate*}
	[label=\roman*)]
  }%
  {\end{enumerate*}}

\newcommand{\code}[1]{\texttt{\small{#1}}}	

\input{terms}

\makeatletter
\def\@copyrightspace{\relax}
\makeatother

%% file: terms.tex
\newcommand{\sys}{Elf}

\newcommand{\OracleOp}{\code{GoldenNN}}
\newcommand{\OptOp}{\code{UniNN}}
\newcommand{\CaptureAll}{\code{CapAll}}
\newcommand{\Oracle}{\code{Oracle}}

\newcommand{\aggwindow}{aggregation window}
\newcommand{\planwindow}{horizon}
\newcommand{\myH}{$\mathscr{H}$}



%% file: abstract.tex
\begin{abstract}

\sys{} is a runtime for an energy-constrained camera to continuously summarize video scenes as approximate object counts. 
\sys{}'s novelty centers on planning the camera's \textit{count actions} under energy constraint.
(1) \sys{} explores the rich action space spanned by the number of sample image frames and the choice of per-frame object counters;
it unifies errors from both sources into one single bounded error. 
(2) To decide count actions at run time, 
\sys{} employs a learning-based planner, jointly optimizing for past and future videos without delaying result materialization. 
Tested with more than 1,000 hours of videos and under realistic energy constraints, 
\sys{} continuously generates object counts within only 11\% of the true counts on average. 
Alongside the counts, \sys{} presents narrow errors shown to be bounded and up to 3.4$\times$ smaller than competitive baselines.
At a higher level, \sys{} makes a case for advancing the geographic frontier of video analytics. 

\end{abstract}

%% file: intro.tex
\section{Introduction}\label{sec:intro}


\paragraph{A case for autonomous cameras}
Today's IoT cameras and their analytics mostly target urban and residential areas with ample resources, notably electricity supply and network bandwidth.
Yet, video analytics has rich opportunities in more diverse environments where cameras are ``off grid'' and connected with highly constrained networks. 
These environments include construction sites, interstate highways, underdeveloped regions, and farms. 
There, cameras must be \textit{autonomous}.
First, they must be energy independent.
Lacking wired power supply, they typically operate on harvested energy, e.g., solar or wind~\cite{solar-camera,solar-AI,ur-solarwind}.
Second, they must be compute independent. 
On low-power wide-area network where bandwidth is low (e.g., tens of Kbps~\cite{lora}) or even intermittent, the cameras must execute video analytics on device and emit only concise summaries to the cloud. 
\sect{bkgnd} will offer more evidence. 

\paragraph{Target query: object counting}
As an initial effort to support analytics on autonomous cameras, 
this work focuses on a key query type: \textit{object counting} with bounded errors.
Inexpensive IoT cameras produce large videos~\cite{wyze-camv2,yi-cam}. 
To extract insights from video scenes, 
a common approach is to summarize a scene with object counts.
This is shown in Figure~\ref{fig:workflow}:
a summary consists of a stream of object counts, one count for each video timespan called an \textit{\aggwindow{}}. 
Object counting is already known vital in urban scenarios;
the use cases include counting customers in retailing stores for better merchandise arrangement~\cite{lipton2015video};
counting audiences in sports events for avoiding crowd-related disasters~\cite{video-public-safety,karpagavalli2013estimating,yogameena2017computer}.
Beyond urban scenarios, object counting further enables rich analytics:
along interstate highways, cameras estimate traffic from vehicle counts, cheaper than deploying inductive loop detector~\cite{inductive-loop,liu2016highway,beymer1997real,naphade2017nvidia};
on a large cattle farm, scattered cameras count cattle and therefore monitor their distribution, more cost-effective than livestock wearables~\cite{IoT-cattle-tracking,digitalanimal-cattle-tracking,postscapes-cattle-tracking};
in the wilderness, cameras count animals to track their behaviors ~\cite{norouzzadeh2018automatically,parham2017animal,van2014nature,hodgson2016precision}.

\input{fig-workflow}


\paragraph{\sys{} and its operation}
This paper presents \sys{}, a runtime for an autonomous camera to produce error-bounded object counts 
with frugal resources, especially limited energy. 
The counts are annotated with confidence intervals (CIs) as shown in Figure~\ref{fig:workflow}.
CI is a common notion in approximate query processing (AQP)~\cite{blinkdb,summarystore,garofalakis2001approximate,olamr,yan2014error,chakrabarti2001approximate}.
A narrower CI suggests higher confidence in the count, hence more useful to users.

\sys{} builds atop periodic energy budgets, an abstraction commonly provided by existing energy-harvesting OSes~\cite{goiri2012greenhadoop,farmbeats}.
As illustrated in Figure~\ref{fig:workflow}, at fixed time intervals of a \textit{\planwindow{}} (e.g., one day), the OS replenishes energy budget to be used by \sys{} in the next \planwindow{}. 
With given energy budget, 
\sys{} periodically wakes up the camera to capture video frames (\circled{1});
it runs neural network (NN) object counters on each captured frame (\circled{2});
by aggregating per-frame object counts, 
\sys{} materializes a per-window aggregated object count for each window (\circled{3});
\sys{} emits the sequence of aggregated counts by uploading them to the cloud either in real time or upon user request. 

\paragraph{The central problem: count planning}
The theme of \sys{} is to produce aggregated object counts with high confidence under limited energy. 
\sys{}'s operational objective for a \planwindow{} \myH{} is to maximize the overall confidence in all the emitted aggregated counts belonging to \myH{} while respecting the energy budget for \myH{}.
This is shown in Figure~\ref{fig:workflow}(a).




The above theme sets \sys{} apart from various visual analytics systems~\cite{noscope,focus,diva,videostorm,chameleon,vstore,wan2019alert}. 
While prior systems focused on per-object results (e.g., accurate object labels), \sys{} takes one step further: 
it directly addresses the need for statistical summaries of videos, optimizing for aggregated counts with narrow errors.  
As a result, while prior systems focused on tradeoffs inherent in vision operators, 
\sys{} exploits higher-order tradeoffs between frame quantities and errors in per-frame counts.
Its design thus has two unique aspects.

\noindent
\textbfit{(1) Per window: characterizing count actions and outcome} \hspace{1mm}
Of an aggregation window $w$, \sys{} navigates in a large space of \textit{count actions}:
it not only chooses the number of sampled frames but also chooses an NN, i.e., a counter, to count objects on individual frames.
This is shown in Figure~\ref{fig:workflow}(b).
For the window $w$, different count actions lead to disparate confidences in the aggregated count as well as disparate energy expenditures.
There is no silver-bullet action.
Notably, a more energy-expensive counter does not necessarily lead to higher confidence:
while per-frame counts have lower errors, \sys{} can afford to process fewer frames, leading to lower confidence in the aggregated count. 


Among all possible count actions for $w$, which ones should \sys{} consider? 
Our answer is an \textit{energy/CI front} -- all the count actions that lead to the narrowest CIs at different energy expenditure levels.
To quantify the energy/CI front, \sys{} integrates as one unified CI two errors: 
i) \sys{} sampling frames rather than processing all possible frames; 
ii) per-frame counts being inexact. 
Prior AQP systems addressed the former~\cite{blinkdb,summarystore,olamr} but never the two integrated to our knowledge.
\sys{} unifies the two errors with novel modeling and approximation, as will be discussed in Section~\ref{sec:design-CI}.



\noindent
\textbfit{(2) Across windows: making joint count decisions on the go} \hspace{1mm}
Of a window $w$, 
the energy/CI front reflects the confidence \textit{return} from  \sys{}'s energy \textit{investment}. 
As we will empirically show, 
windows often have different energy/CI fronts.
As a result, investing the same amount of energy on different windows yields disparate CIs in their respective object counts. 
As \sys{}'s objective is to maximize the overall confidence for a \planwindow{}, 
it must decide heterogeneous count actions for windows jointly, 
e.g., to invest more energy on windows where CI width sees more substantial reduction. 
 


To do so, 
\sys{} addresses a dilemma between 
the needs for \textit{global knowledge} and for \textit{timely decisions}: 
i) to decide the optimal count action for a window $w$, 
\sys{} must compare the energy/CI fronts of all windows in the \planwindow{}; 
ii) to timely materialize the aggregated count of $w$, 
\sys{} must decide the count action for $w$ without seeing \textit{future} windows.
To this end, 
\sys{} \textit{predicts} the action count for each upcoming window 
based on the observation of past windows. 
Based on reinforcement learning, \sys{} mimics what an oracle planner, which (impractically) knows a \planwindow{}'s 
all past and future windows, \textit{would} decide. 

\sys{} runs on off-the-shelf embedded hardware.
Our tests on over 1,000-hour real-world videos demonstrate \sys{}'s efficacy: 
with the energy level of a small solar panel, 
\sys{} continuously produces aggregated object counts that are only within 11\% of the truth counts.
At 95\% confidence level, the CI widths are as narrow as 17\% on average.
We make the following contributions. 

\begin{myitemize}
\item 

\textit{The design space:}
We explore object counting on autonomous, energy-constrained cameras. 
We identify the central design problem: characterizing count actions and choosing them at run time. 

\item 
\textit{The problem formulation:}
From the large space of count actions, 
we formulate an energy/CI front as the viable actions a camera should consider. 
To quantify the front,
we propose novel techniques for unifying errors from multiple sources into a single CI.

\item 
\textit{The runtime mechanisms:}
To plan count actions at run time, 
we design a novel, learning-based planner, which mimics the decisions made by an oracle under the same observation of past videos. 

\item 
\textit{The implementation:}
We report a prototype \sys{}. 
Running on commodity hardware and with energy from a small solar panel, 
\sys{} continuously produces video summaries with high confidence. 



\end{myitemize}

To our knowledge, \sys{} is the first software system executing video object counting under energy constraint. 
Our experiences make a case for advancing the geographic frontier of video analytics. 

%% file: fig-workflow.tex

\begin{figure}[t]
	\centering
	\includegraphics[width=0.49\textwidth]{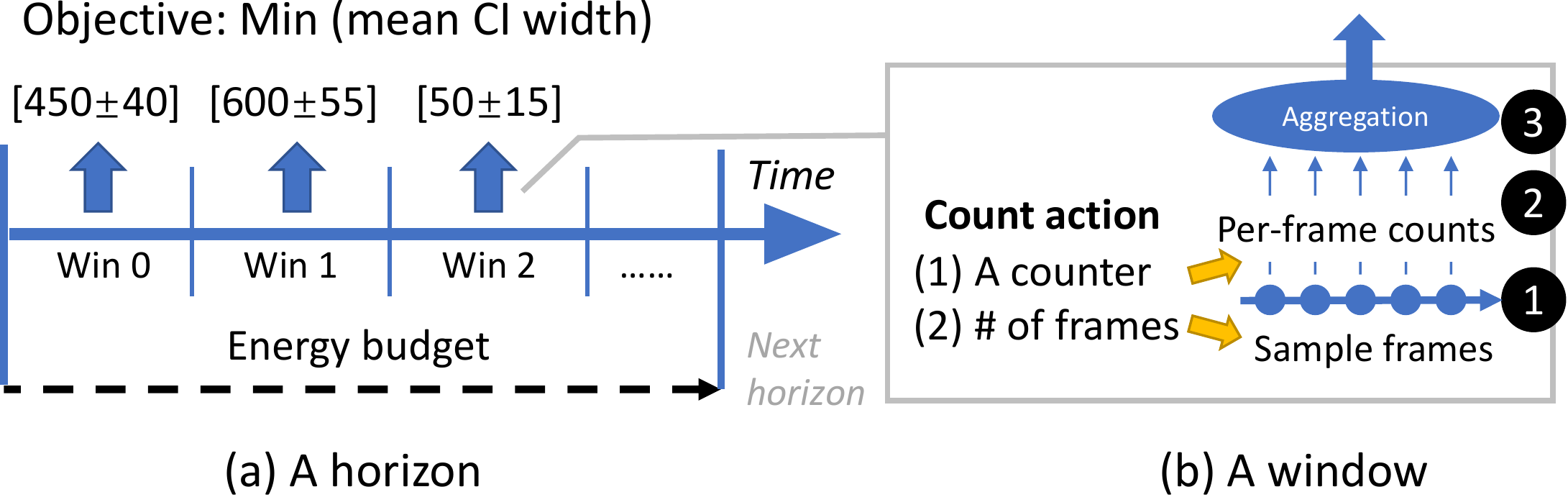}
	\caption{The target analytics and its execution. 
	For simplicity, only counts for one object class are shown.}
	\label{fig:workflow}
\end{figure}


%% file: bkng.tex

\section{Background}\label{sec:bkgnd}

\subsection{Autonomous Cameras}
\label{sec:bkgnd-autonomous}

\paragraph{Compute: commodity SoCs}
Similar to commodity IoT cameras~\cite{yi-cam,wyze-camv2}, autonomous cameras incorporate embedded application processors, e.g., those with Armv7a or MIPS32 cores at a few GHz and a few GBs of memory. 
We do not assume special-purpose hardware for vision~\cite{solar-AI,redeye}, as their wide adoption remains to be seen.
The cameras run commodity OSes like Linux~\cite{gopro-os}, which supports POSIX apps and frameworks, e.g., TensorFlow.

\paragraph{Network}
In off-grid scenarios we target, 
low-power wide-area networks (LPWAN) emerge as the norm~\cite{lpwan-market}.
One popular standard is LoRaWAN~\cite{lora}: 
engineered for long-range, low-power communications, 
it mandates deeply duty-cycled (<1\%) transmissions with data rate as low as a few Kbps~\cite{limits-lorawan}. 
Such networks motivate cameras to upload concise video summaries only, e.g., object counts over time windows. 
By contrast, 
uploading image frames (720P), even one in every one minute (inadequate for deriving useful object counts as our evaluation will show) 
would consume at least 40 Kbps per camera~\cite{limits-lorawan}, unsustainable on LPWAN.

\paragraph{Energy source}
A key parameter of our system is the typical range of energy budgets. 
We use as a reference energy source a small (20cm $\times$ 20cm) solar panel backed by a rechargeable energy buffer.
Such an energy harvester costs as low as \$30,
commonly seen in embedded prototypes and production~\cite{solar-traffic-lights,UR-SolarCap}.

In two ways, we estimate the daily energy budget available from such an energy source. 
First, we measure the energy harvested by our small solar panel in the Midwestern US as 12.6 -- 20.5 Wh/day. 
Second, we follow the prior studies on solar-powered embedded systems~\cite{UR-SolarCap} and their typical parameters:
a solar panel rated at 5 W, 
an energy buffer as a supercapacitor of 60Wh, 
and solar irradiance of 2490 -- 9629 Wh/$m^2$/day (major US cities, according to the National Solar Radiation Database~\cite{solar-radiation-database}). 
The energy available for use is 12.45 -- 48.15 Wh/day.
We conclude to design and evaluate \sys{} with energy budgets in the range of 10 -- 30 Wh/day. 

\paragraph{Energy budgeting}
At run time, the camera OS allocates an energy budget prior to each \planwindow{}. 
OS energy budgeting has been well studied~\cite{farmbeats,goiri2012greenhadoop,goiri2014designing,singh2013yank} and is orthogonal to this work. 
In a nutshell, 
it sets the \planwindow{} length to reflect periodic/temporal energy availability, e.g., one day; 
it allocates energy budgets in order to stay energy neutral given future energy availability 
(e.g., sunlight in the next few days). 
As such, the OS may set different energy budgets for different \planwindow{}s.
With the OS-level energy budgeting, \sys{} neither has to be the only application running on an autonomous camera nor has to run one query only.

\subsection{Video Summary via Object Counting}
\label{sec:bkgnd-query}

At camera deployment time or run time,
a user specifies her analytics as a continuous query $ \langle \tau,\alpha, K \rangle$: 

\begin{myitemize}
\item 
\textit{Aggregation window length} ($\tau$) defines the temporal granularity, e.g., 30 mins, at which object counts are aggregated. 

\item 
\textit{Confidence level} ($\alpha$) specifies the desired probability of
the query answer covering the ground truth count, e.g., 95\%.
Based on the desired probability \sys{} generates confidence interval (CI), a bounded error widely adopted in analytics systems~\cite{blinkdb,summarystore,garofalakis2001approximate,
olamr,yan2014error,chakrabarti2001approximate}.

\item 
\textit{Object classes} ($K$, optional). 
By default, the camera counts all the object classes recognizable by modern vision object detectors (e.g., 80 classes by YOLOv3 trained on COCO~\cite{coco}). 
The user may narrow down the counted classes to a subset $C$, e.g., cars and humans. 

\end{myitemize}

\sys{} answers a query with a stream of aggregated counts, 
$\{ [ \mu_{i,j} \pm \delta_{i,j} ] \}$. 
In the sequence, each tuple corresponds to one window; 
$\mu_{i,j}$ is the output aggregated count of object class $j$ in window $i$; 
$\delta_{i,j}$ is the CI width, 
which covers the true count with $\alpha$\ probability. 
For example, 
under $\alpha=95\%$ when \sys{} emits a count $1000 \pm 100$,
it indicates that the probability for the true count to fall in (900, 1100) is 95\%.
A smaller $\delta$ indicates higher confidence in the count, making it more useful.
Note that \aggwindow{}s and CI are well-known concepts used by popular analytics systems~ \cite{summarystore,blinkdb}.
Through further analysis, users may compose object counts accompanied by CIs  into higher-level statistical summaries that describe trends in long videos, e.g., ``there is a 90\% probability that cars crossing this intersection have doubled in the past week''.

\subsection{Object Counters on Individual Frames}\label{sec:bkng-NN}
As described in $\S$\ref{sec:intro},
to derive the object count for a window, 
\sys{} first counts objects on individual frames sampled from that window.
To count per-frame objects,
we follow a common approach of executing object detectors~\cite{wei2019city,tf-vehicle-det,biswas2017automatic,furuya2014road}.
The state-of-the-art object detectors are neural networks (NN). 
We choose a set of generic, popular detectors as listed in Table~\ref{tab:ops}.
Note that: 
i) We are aware of object detectors specialized for particular videos~\cite{noscope} which
are compatible with our design. 
We motivate and validate our design with generic, well-known detectors for ease of experiments and result reproducibility.
ii) We are aware of recent work using NN to emit counts without first detecting objects~\cite{blazeit}; 
we dismiss such an approach due to its tedious per-video, per-class training and much lower accuracy than object detectors observed in our experiments.

The wide selection of NN counters offers diverse energy/error tradeoffs at the frame level. 
This is crucial, as no single counter can yield the narrowest CIs for all the windows under a given energy budget. 
For instance, \code{YOLOv3}, an expensive counter, incurs low error in per-frame counts;
however, with its high energy cost (e.g., $\sim$50J per frame on an Arm device, see \S\ref{sec:impl}) the system can only afford processing one frame in every 2 minutes (with a 10Wh/day budget), which eventually leads to inferior CIs.
We will show more evidence in $\S$\ref{sec:overview}.

%% file: overview.tex
\section{The \sys{} Design}
\label{sec:overview}



\input{fig-system}

Figure~\ref{fig:system-model} shows the architecture of \sys{}.
A user installs a query to the camera either at the deployment time or over the air later.
\sys{} plans its execution in the scope of a \planwindow{} \myH{}.






\subsection{System Operation}\label{sec:operation}

\paragraph{Energy expenditure} 
\sys{} executes the query by respecting the received energy budget for the \planwindow{} \myH{}.  
It spends the energy on the following activities:
(1) $E_{cap}$ for capturing frames; 
(2) $E_{count}$ for executing counters on frames; 
(3) $E_{agg}$ for deriving aggregated counts; 
(4) $E_{upload}$ for uploading the aggregated counts. 
The first two activities dominate the \sys{}'s energy consumption (>99.9\% as we measured):
in each window (e.g., 30 min), 
\sys{} acquires tens of MB pixels from image sensors and runs several trillions of FLOPs. 
Activities (1) (2) hence will be our focus. 
By comparison, the latter two consume negligible energy:
(3) only incurs hundreds of arithmetic operations per window and
(4) only uploads concise numerical counts no more than a few hundred bytes per window. 



\paragraph{Count action} 
For each window belonging to \planwindow{} \myH{},
\sys{} picks the number of frames to sample and an object counter.  
For the window, the action determines the confidence in the window's aggregated count and energy expenditure (including both $E_{cap}$ and $E_{count}$). 
Note that \sys{} must run the same counter on all frame samples from a window in order to integrate per-frame count error in a tractable way (see $\S$\ref{sec:design-CI} for details); 
from a window, it must draw frame samples uniformly in time to avoid sampling bias~\cite{intro-stats}. 


\paragraph{Objective: overall confidence as mean CI width}
Operating under the energy budget for \myH{}, 
\sys{} seeks to maximize the overall confidence for \myH{}. 
Our implementation defines the overall confidence as reciprocal to the mean of CI widths of the counts from all the windows in \myH{}.  
Mean CI width is commonly used to characterize the overall confidence in a set of aggregates~\cite{summarystore}.
\sys{} is also compatible with alternative overall confidence metrics, such as median and minmax of CI widths of all windows in \myH{}. 

\subsection{Count Action \& Outcome}
\label{sec:overview:problem}


We next dive into the relation between the count action and the outcome. 
To simplify discussion, we focus on counting for a single object class (i.e., one aggregated count per window);
yet the discussion applies to multiple classes if we consider the mean CI width over all the counted classes.

For a window $w$: 

\begin{footnotesize}


$\underbrace{\langle N_w,C_w \rangle}_\text{Action} \rightarrow \underbrace{<E_w,\delta_w>}_\text{Outcome}$
\end{footnotesize}

where $N_w$ is the counter choice and $C_w$ is the number of frames to sample; 
$E_w$ is the energy spent on the window (including both $E_{cap}$ and $E_{count}$) 
and $\delta_w$ is the CI width for the aggregated count of the window. 
For instance, \sys{} may execute an \textit{expensive} counter (e.g., a deeper NN) to produce \textit{more} exact per-frame counts with \textit{higher} per-frame energy; 
or a \textit{cheaper} counter (e.g., a shallower NN) produces \textit{less} exact counts with \textit{lower} energy.
Note that while varying sample quantity is widely exploited by prior AQP systems, varying per-sample errors (in our case, through the counter choice) is a less explored opportunity as enabled by NNs. 

\paragraph{The action/outcome plot}
All possible $N \times C$ combinations 
present rich actions available to \sys{}. 
They are visualized in action/outcome plots in Figure~\ref{fig:opt-op}. 
On one plot, picking one counter (a curve in the plot) and the number of sample frames (a point on the curve with the number annotated), 
\sys{} generates an aggregated count with specific CI width (Y-axis) at energy consumption (X-axis).
Along a given curve, as sample number increases, CI width narrows, according to sampling theory~\cite{intro-stats}, and energy consumption increases. 
All such plots for all windows in \myH{} form the basis for \sys{} to make count decisions. 

\paragraph{How to derive an action/outcome plot?}
For a specific window $w$, the plot is determined by
(1) the true object count in $w$; 
(2) the count variation during the course of $w$; 
(3) the per-frame count errors of candidate counters. 
$\S$\ref{sec:design-CI} will present quantification details.

\paragraph{The design implications} are as follows:

\input{fig-opt-op}

%

\begin{myitemize}

\item 
\textbf{No counter is silver bullet}.
For a given window, no single counter 
(e.g. the most expensive or the cheapest ones) 
\textit{always} results in the narrowest CI. 
As exemplified in Figure~\ref{fig:opt-op} (left): 
when the window's energy expenditure is low ($<$0.5 kJ), cheap counters result in narrower CIs; 
as the energy expenditure increases, expensive counters start to excel. 
This is because with less energy the CI is primarily bottlenecked by the inadequacy of sample frames, and cheap counters allow more sample frames. 
With abundant energy, even running more expensive counters \sys{} can afford adequate frames;
hence, the CI starts bottleneck at the errors in per-frame counts.

\item 
\textbf{Operate on the energy/CI front only.} 
On an action/outcome plot,
the bottom-left segments from multiple curves constitute an energy/CI front 
as highlighted in Figure~\ref{fig:opt-op}.
The front contains all the ``optimal'' count actions that lead to the minimum CI width (Y-axis) with different amounts of energy spent on the window (X-axis). 
The implications are that:
i) \sys{} should always operate on the energy/CI front, picking a point of count action from the front according to how much energy it is willing to spend on that window. 
ii) As \sys{} considers spending additional energy on a window,
the gradient of the front indicates \textit{prospect} of confidence gain, i.e., the rate of CI width reduction in response to additional energy investment.

\item
\textbf{Make joint decisions across windows.}
The energy/CI fronts vary across windows, as exemplified by a comparison of Figure~\ref{fig:opt-op} (left) and (right). 
This means the same amount of energy expenditure on different windows will result in different CI widths. 
For example, increasing the energy from 2.0 kJ to 3.0 kJ would reduce the CI width by 26 in Figure~\ref{fig:opt-op} (right) but only 9 in (left). 
Intuitively, windows with higher object counts and higher variations (e.g. video clips during rush hours) would see higher CI reduction than otherwise (e.g., midnight). 
Since the objective of \sys{} is to minimize the mean CI width across a \planwindow{} ($\S$\ref{sec:operation}), 
it shall decide count actions across windows \textit{jointly}, resulting in heterogeneous actions and outcomes.
As will be shown in the evaluation, the resultant energy expenditures across windows could differ by 9x. 

\revise{
\paragraph{Putting it together}
The above observations suggest an energy planning strategy as follows.
Based on energy/CI fronts for individual windows, 
the system iteratively invests energy slices to the window that sees the highest prospect of confidence gain.
For that invested window, the system picks the optimal counter and frame count according to the window's energy/CI front. 
Intuitively, the system is inclined to invest more energy on windows with higher object counts and variations; the system is therefore more likely to sample more frames and pick more expensive counters -- as energy budget permits -- on such windows. 
This strategy is the basis of the oracle planner in $\S$\ref{sec:plan:oracle}.
}


\end{myitemize}

%% file: fig-system.tex

\begin{figure}
	\centering					
	\includegraphics[width=0.42\textwidth]{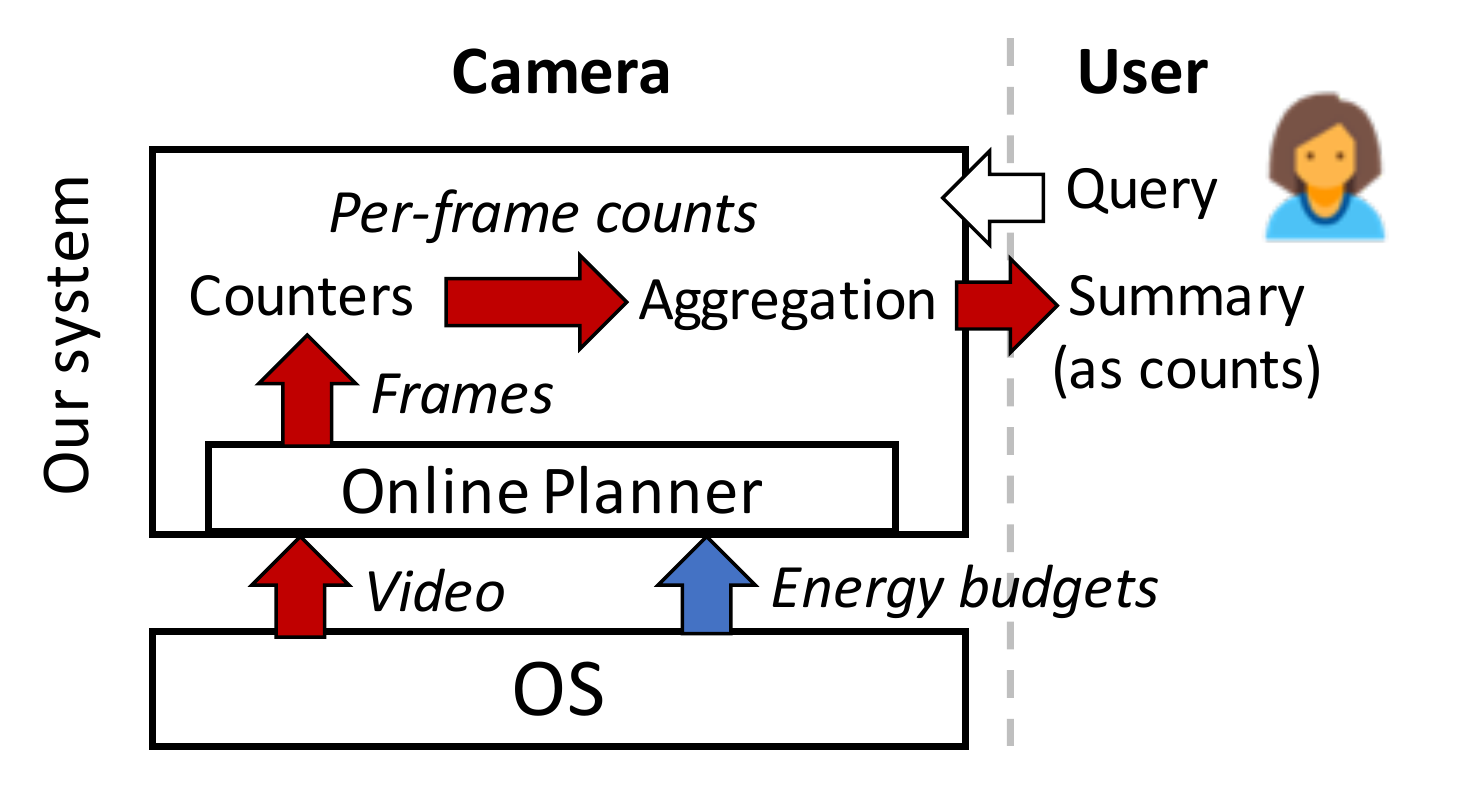}
	\caption{The \sys{} architecture}
	\label{fig:system-model}
\end{figure}

%% file: fig-opt-op.tex
\begin{figure}
	\centering					
	\includegraphics[width=0.48\textwidth]{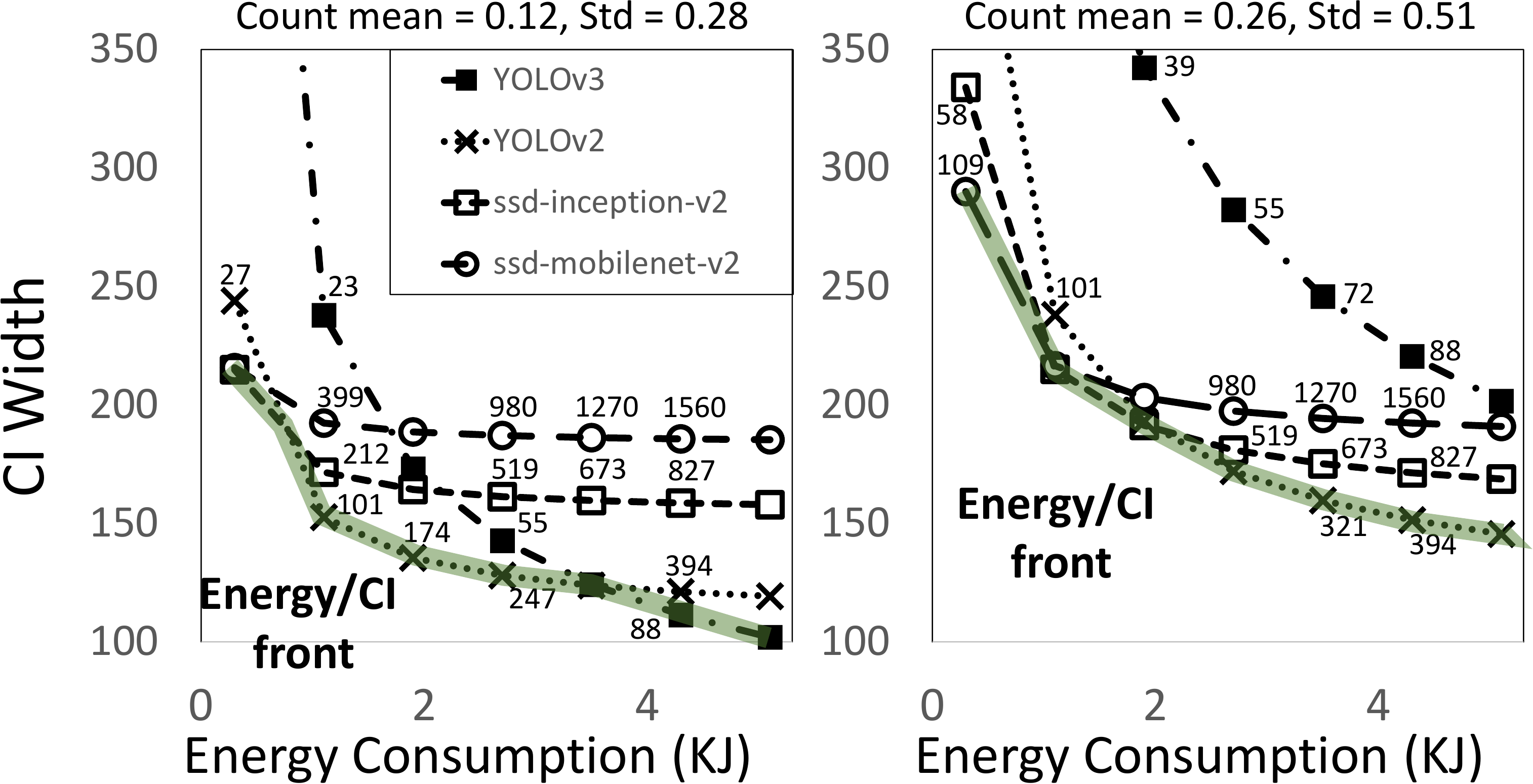}
	\caption{
	A comparison of action/outcome plots for two example windows, 
	showing their disparate outcomes and energy/CI fronts. 
On each plot: 
X-axis shows the window's system energy expenditure ($E_{cap}$ and $E_{count}$); 
Y-axis shows the CI width of the aggregated count for that window; 
each curve corresponds to a counter choice, on which each point corresponds to an amount of frames to sample (annotated along curves). 	
}
\label{fig:opt-op}
\end{figure}

%% file: design.tex
\input{design-plan}
\input{design-CI}

%% file: design-plan.tex
\section{Count Planning}
\label{sec:plan}


\paragraph{Challenge}
To plan count actions towards the objective described in $\S$\ref{sec:operation}, 
\sys{} shall address a twist of two needs:

\begin{myitemize}
\item The need for \textbf{global} knowledge. 
To optimize for the whole \planwindow{}  \myH{},
\sys{} shall make joint count decisions across windows (\S\ref{sec:overview:problem}), 
i.e. comparing the energy/CI fronts of all windows, 
which reflect their confidence gains from potential energy investment.
Yet, not until the end of \myH{} can \sys{} estimate the fronts of all windows belonging to  \myH{},
as \sys{} needs to see their respective object counts and count variation (see $\S$\ref{sec:design-CI} for details). 



\item 
The need for \textbf{on-the-go} decisions. 
Deferring count actions to the end of \myH{} (as indicated above) raises two problems. 
i) 
Stale results:
\sys{} will not be able to emit the counts for all windows until the end of  \myH{}. As  \myH{} may span hours or days, doing so prevents users from observing fresh object counts. 
ii) 
Excessive capture:
Prior to each window, \sys{} must decide the number of frames to \textit{capture}, which limits the number of frames \sys{} can \textit{process} later. 
Deciding count actions at the end of  \myH{} forces \sys{} to play safe, capturing excessive frames for each window.
Besides the two, 
holding captured frames until the end of  \myH{} increases camera storage pressure and risk of privacy breach.

\end{myitemize}

\paragraph{Approach overview}
\sys{} addresses both needs above with an online planner. 
The key idea is to make online decisions by mimicking what an oracle planner would do. 
Specifically, 
the oracle planner works \textit{offline}: 
with full knowledge of a video, it decides what count actions \textit{should have been} for windows in the video by considering the energy/CI front of all the windows jointly. 
Trained with the oracle decisions and true object counts from the videos,
the online planner makes decisions that the oracle \textit{would} make with a similar observation of recent windows. 

\input{design-onephase}
\input{design-RL}

\input{design-splitphase}

%% file: design-onephase.tex

\subsection{The Oracle Planner (Offline)}
\label{sec:plan:oracle}
The oracle planner works based on impractical assumptions:
i) it knows the energy/CI fronts of all windows and 
ii) the amount of captured frames exactly matches the amount needed in processing.
The oracle plans count actions by solving an energy allocation problem:
it iteratively allocates energy slices to the window that exhibits the highest CI width reduction. 
More specifically, for a \planwindow{}: 

\begin{myitemize}

\item \textit{Initialization:} The oracle planner dispatches energy to each \aggwindow{} so that each window has a minimum number $M$ of frame samples. 
This is because in statistics, an estimation from sampling is only regarded meaningful when the sample size is sufficiently large.
We pick $M=30$ by following common practice in statistics~\cite{kar201330}.
The oracle picks the cheapest NN counter for each \aggwindow{} in order to start from the lowest possible energy consumption.

\item \textit{Iteration:} 
The oracle repeatedly allocates a small, fixed amount of energy to individual windows. 
To pick the next window $W$ for receiving an energy slice, the oracle examines the energy/CI fronts of all windows, 
selecting $W$ to be the one having the highest gradient $|\frac{\partial CI}{\partial E}|$ at the current operating point $\langle$ energy, CI width $\rangle$. 
By allocating the energy slice, the oracle updates the count action for $W$ and slides $W$'s operating point along its energy/CI front.


\item \textit{Terminate:} The oracle planner stops when it uses up the energy budget. 
At this moment, the NN counters and numbers of frames for all windows are the final count decisions.
\end{myitemize}


In evaluation, we consider the oracle as the upper bound of the overall confidence achievable by \sys{}. 

%% file: design-RL.tex
\subsection{The Learning-based Planner (Online)}
\label{sec:design:rl}

\input{fig-rl}

\paragraph{Overview}
Prior to \sys{} deployment, we run the oracle planner offline on sample video footage from the target camera; in our implementation we use 3-day video footage. 
Using the video and the oracle decisions as the training dataset, we train the online planner. 
Deployed on the camera, the online planner continuously outputs count decisions
based on the object counts and count variation that \sys{} emits recently, including those from recent windows and the windows at similar times in recent days.




\paragraph{Rationale: Why could an online planner work?}
An online planner acts only based on its observation of the past;
yet, as we will demonstrate in $\S$\ref{sec:eval}, it can output decisions closely matching the oracle that knows both the past and the future. 
We attribute the efficacy to the temporal correlation among object counts and count variations in a video feed, a pervasive video characteristic. 
For instance, the car count in 9 AM -- 9:30 AM correlates to the counts from half-hour windows prior to 9 AM of the same day and to the count from 9 AM -- 9:30 AM of prior days. 
While the correlation is still not deterministic enough for \sys{} to directly predict counts as analytics results,
the correlation provides sufficient \textit{hints} for \sys{} to plan count actions and manage energy.

Intuitively, the count action picked by oracle for a window $W$ primarily depends on the relative significance of $W$'s object count and variation with regard to those of other windows. 
Due to the temporal correlation discussed above, both information is encoded in the sequence of past object counts and count variations prior to $W$. 
As such, the online planner can use the past sequence to predict the oracle's decision for $W$.

\paragraph{Why reinforcement learning (RL)?}
Essentially, the online planner takes sequential actions to optimize a long-term objective (i.e. mean CI width).
This pattern well suits RL, a general framework for sequential decision making.
In RL, an ``agent'' (e.g., our planner) interacts with its environment (e.g., the energy budget and all energy/CI fronts) in discrete time steps to maximize its cumulative reward. 
If the environment is fully observable to the agent, 
the agent takes an action based on the current environment state at each time step.
The action takes the environment to a new state, and the agent receives a reward accordingly. 
If the environment is only partially observable (e.g., our planner only knows past, but not future, energy/CI fronts), the agent takes an action based its observation $o_t$ of the environment. 

To apply RL, we face the following challenges. 

\begin{myitemize}
	\item
	(C1) Long delay in rewarding.
	The count actions of all windows jointly decide the mean CI width for the whole \planwindow{} \myH{}.  
	Hence, not until the end of \myH{} can \sys{} evaluate its past count actions and assign reward/penalty accordingly.
	This makes RL training difficult. 
	
	\item 
	(C2) Hard constraint on the accumulative outcome. 
	Of our online planner, the total energy expenditure across multiple steps (i.e. windows) should respect a constraint -- the energy budget for \myH{}.
	Yet, the standard RL lacks mechanism to enforce such a constraint to our knowledge. 
	
\end{myitemize}


	We address (C1) to reward the planner in training timely and frequently -- at every time step:
	rather than reasoning about the long-term impact of the planner's decision, 
	we consider how much the decision deviates from the oracle's decision. 
	We address (C2) by making the planner \textit{implicitly} learn to respect an energy budget, as we train the planner to follow the decisions made by the oracle that operates under the same energy budget. 
	To handle the unlikely events that the planner burns out the budget before a \planwindow{}'s end,
	we devise a backstop mechanism to be described later.

\paragraph{RL formulation}
As illustrated in Figure~\ref{fig:rl}, we formulate:

\begin{myitemize}
	
	\item 
	\textit{A time step} is an \aggwindow.
	
	\item 
	\textit{The observation} vector consists of the object counts and count variation of the most recent $M$ windows and the same-time windows in the recent $N$ days. 
	Our experiment empirically chooses $M=4$ and $N=1$.
			
	\item 
	\textit{The agents} are two that receive the same observation and 
	pick the number of sample frames and the counter, respectively. 
	We instantiate the two as a regression agent and a classification agent. 
	Both agents are NNs with the same multi-layer perceptron (MLP) architecture. 
	The NNs are small: the input layer has only 10 input units; 
	both NNs have two hidden layers each with 64 hidden neurons only. 
	Each NN has less than 5K parameters and 55KB in size.
		
	\item 
	\textit{The reward/penalty} is an agent's decision deviation from the oracle's decision. 
	For the regression agent, its reward function is:
		$ r_{t,reg} = - | N_{t,agent} - N_{t,oracle} | $
	where $N_{t,agent}$ and $N_{t,oracle}$ are the frame amounts from the agent and from the oracle, respectively.
	For the classification agent, the reward function is:
	$\nonumber
	r_{t,cls} = 1 \textit{ if } (C_{t,agent} = C_{t,oracle}) \textit{ otherwise } 0$, 
	where $C_{t,agent}$ and $C_{t,oracle}$ are the NN counters chosen by the agent and the oracle planner respectively.
	
	
	\item Respecting \textit{energy budgets}. 
	To train the RL agents for operating on a variety of energy budgets that the agents may receive from the OS at run time, 
	we discretize the target energy budget range into multiple levels (e.g., 10--30 Wh/day at 5 Wh/day steps) and train a pair of RL agents for each energy level.
\end{myitemize}

\paragraph{Offline training \& cost}
We follow a standard approach: 
we use the Actor-Critic framework~\cite{actor-critic}
and A2C~\cite{a2c}, a synchronous, deterministic training algorithm as a variant of A3C~\cite{a3c}. 



\revise{
Before deploying RL agents in real-world systems, we first run the oracle planner on a video segment (3 days in our experiment) to collect training data and train the RL agents offline.
Our experience shows modest training effort in general. 
For most video scenes under test (listed in Table~\ref{tab:dataset}), 
we find a 3-day video sufficient for training. 
The training overhead primarily comes from: 
i) running the oracle planner, including deriving the energy/CI fronts by testing all candidate counters on the videos; 
ii) training the RL agents. 
The former takes a few hours on a commodity GPU workstation with one Nvidia Quadro 6000 and can be further accelerated by additional GPUs or TPU; 
the latter takes as low as tens of minutes on the same workstation. 

After deployment, we expect the trained RL agents to operate for a long period of time autonomously. Our experiments show their stable accuracy  over our longest video lasting 2 weeks.
In real-world deployment, we expect to only retrain the RL agents in case of substantial changes in video scene, e.g. due to cameras being relocated or small scene changes accumulated over time. 
Users may initiate retraining based on their knowledge on camera deployment, or as preventive maintenance. 
Similar to the initial training, users would need to retrieve a recent video segment from the camera, run the oracle over the video on their development machines, and update the RL agents on the camera. 
}

\paragraph{Online prediction \& cost}
Once trained, the online planner is deployed as part of \sys{} on camera.
As RL is incapable of guaranteeing perfect decisions, 
what if the planner mispredicts (despite unlikely)? 
In particular, burning out energy before a \planwindow{} ends would 
leave no energy for the remaining windows and therefore no counts produced for them. 
To this end, \sys{} incorporates a backstop mechanism alongside the planner.

Over the course of a \planwindow{}, \sys{} monitors the energy balance, e.g., the remainder from the budget. 
When the balance drops down to the ``bare minimum'', i.e., the amount needed by the remainder windows to run the cheapest counter with the smallest frame count needed to be statistically significant (e.g., 30), \sys{} bypasses the planner for the remaining windows and follows the minimum count actions.
If the online planner acts conservatively and has not used up the energy when a \planwindow{} ends, \sys{} returns the unused energy to the OS.

We encourage \sys{}'s conservative energy expenditures through tuning reward functions for our RL agents. 
The rationale is that we prefer (slight) budget underutilization to early burn out: 
while the former only sees minor loss in the overall confidence, 
the latter results in significant CI widths for a series of windows. 

The planner itself incurs negligible energy overhead. 
For each window, it performs around 8K multiply-adds operations. 
By contrast, even the cheapest NN counter in our selection performs 0.8 billion multiply-adds operations \textit{per frame}. 
We estimate the planner's energy consumption is at least four orders of magnitude smaller than the per-frame counter.

%% file: fig-rl.tex
\begin{figure}[t]
	\centering
	\includegraphics[width=0.45\textwidth]{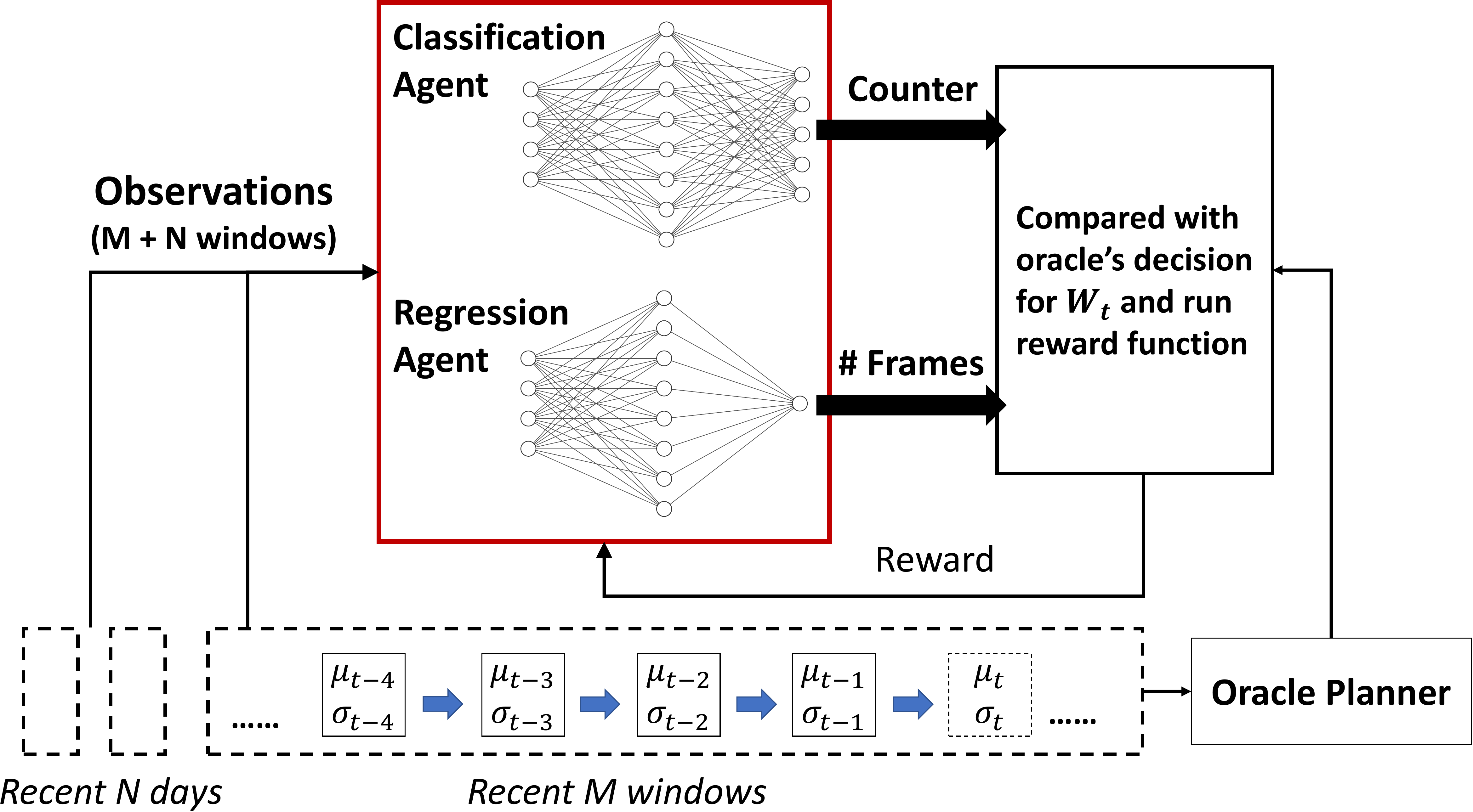}
	\caption{Training RL agents for online count decision. $\mu$ and $\sigma$ represent the object count and count variation (observations) for each window.}
	\label{fig:rl}
\end{figure}

%% file: design-CI.tex

\setlength{\belowdisplayskip}{5pt} \setlength{\belowdisplayshortskip}{1pt}
\setlength{\abovedisplayskip}{5pt} \setlength{\abovedisplayshortskip}{1pt}

\section{Integrating Multi-Source Errors}\label{sec:design-CI}

\input{tab-notations}

\input{fig-CI-workflow}


\paragraph{The problem}
Given a set of per-frame counts, 
\sys{} needs to derive an aggregated count accompanied by a \textit{single} CI -- for both planning (i.e., constructing energy/CI fronts for all windows) and for materializing aggregated counts. 

Table~\ref{tab:notations} summarizes the notations used below.
Our \textit{input} is a series of inexact per-frame counts ($x=x_1...x_n$) as observed by one given NN counter (\circled{1}).
Our \textit{output} is a CI as an estimation for $\mu$, i.e., the mean of \textit{exact} per-frame counts on \textit{all} the frames (\circled{5}).


\paragraph{Approach overview}
To model the random variable $\mu$, 
we first model a random variable $\mu_x$, the mean of \textit{inexact} per-frame counts on \textit{all} the frames; 
we derive $\mu_x$ from $x$, the observed \textit{inexact} counts on \textit{sampled} frames (\circled{2}).
We then model the deviation between $\mu_x$ and $\mu$ as caused by errors in per-frame counts (\circled{3}). 
Eliminating the deviation from $\mu_x$, we derive the distribution of $\mu$, 
from which we estimate its CI via approximation (\circled{4}).

We demonstrate the validity of the resultant CIs with experiments on real-world videos ($\S$\ref{sec:eval-perf}).

\paragraph{Modeling $\mu_x$}
The theory of statistics gives us:
\begin{equation}\label{eq:1}
\frac{\bar{x}-\mu_x}{S/\sqrt{n}} \thicksim T_{n-1}
\end{equation}
where $\bar{x}$ and $S$ are the mean and standard deviation of the observed inexact counts ($x$), 
and $T_{n-1}$ is the well-known \textit{t}-distribution with \textit{n - 1} degrees of freedom~\cite{intro-stats}.
This holds regardless of $x$'s actual distribution, e.g., normal or Poisson, as long as the sample population is sufficiently large (e.g., 30 which is also the minimal sampling number for each \aggwindow{} used by \sys), according to the central limit theorem~\cite{intro-stats-2}. 

\paragraph{Modeling the deviation between $\mu_x$ and $\mu$}
The deviation is because $\mu_x$ incorporates errors in per-frame counts that $\mu$ does not incorporate. 
We model the deviation with a heuristics: the deviation has positive correlation with $\mu$. The rationale is that an NN counter is likely to incur more false positives/negatives on video frames with more objects. 
Based on our experiments, 
when the absolute value of $\mu$ exceeds a threshold $\theta$ (as we will empirically determine), the distribution of deviation $E'$ is best modeled as the ratio between $\mu$ and $\mu_x$ (i.e., $\mu / \mu_x \thicksim E'$); 
when $\mu$ is small, the distribution $E''$ is best modeled as a linear offset between the mean of true counts and that of observed counts (i.e., $\mu - \mu_x \thicksim E''$).
Notably, $E'$ and $E''$ cannot be unified as one distribution. 
For instance, a small $\mu$ results in a large ratio;
including those outliers in the overall error distribution results in high distribution variance as observed from real videos.
As \sys{} cannot directly observe $\mu$ at run time, it uses $\bar{x}$ to estimate if $\mu$ exceeds the threshold $\theta$.




\paragraph{Determining the parameters of deviations} 
The means and standard deviations of $E'$ and $E''$ are crucial to modeling $\mu$, which we obtained through offline profiling: 
for each camera, profiling once at the camera's deployment time, and profiling again if the video scene changes significantly, e.g., the camera being relocated. 
This is based on twofold observation below. 

\begin{myitemize}
\item 
The distributions of deviation are stable over time, with a given NN counter and a given video feed. 
As an example, we measured the distributions with two counters, \code{YOLOv2} and \code{ssd mobilenet-v2} and video segments from Jackson; 
we tested two week-long video segments that are one month apart. 
The Bhattacharyya coefficient~\cite{bhattacharyya1943measure}
between the two videos are 0.93 and 0.86 for the two NN counters, respectively.
\item The distributions are independent of the observed object counts $x$, 
as we confirmed with chi-square test~\cite{intro-stats} on our video datasets. 
Given such independence, we can integrate $\mu_x$ and $E$ in order to estimate the distribution of $\mu$, as shown below.
\end{myitemize}

Note that across different NN counters the above distributions of deviation ($E$) are often disparate, even on the same video feed. 
This explains why we only use one NN counter in one \aggwindow{}.

\paragraph{Modeling $\mu$}
We use $V'$ and $V''$ to denote the distributions of $\mu$:

\begin{equation}\label{eq:3}
\mu =
   \begin{cases}
     (\bar{x} + S/\sqrt{n} \cdot t) \times e' \thicksim V' & \quad \text{if } \bar{x} > \theta\\
     (\bar{x} + S/\sqrt{n} \cdot t) + e''  \thicksim V'' & \quad \text{if } \bar{x} \leq \theta
   \end{cases}\\
\end{equation}
\[
   where: t \thicksim T_{n-1}, e' \thicksim E', e'' \thicksim E''
\]

Hence, the CI of $\mu$, denoted as $v'_{\alpha}$ and $v''_{\alpha}$,  is as follows:
\begin{equation}\label{eq:4}
CI =
\begin{cases}
[\bar{x} \times \mu({e'}) \pm v'_{\alpha}] & \quad \text{if } \bar{x} > \theta\\
[\bar{x} + \mu({e''}) \pm v''_{\alpha}] & \quad \text{if } \bar{x} \leq \theta
\end{cases}\\
\end{equation}
\begin{flalign*}
where \quad & {P(|V'-\bar{x}\times\mu({e'})|\leq v'_{\alpha})=\alpha \%}\\
& {P(|V''-\bar{x} - \mu({e''})|\leq v''_{\alpha})=\alpha \%}
\end{flalign*}

Typically, the CI widths are derived through Monte Carlo simulation~\cite{monte-carlo}: randomly picking the same number of samples from \textit{t}-distribution and NN counter's error distribution ($E',E''$) respectively, and combining them as Equation~\ref{eq:3}.
The results are expected to follow the distribution of $\mu$, from which one  gets the CI.


\paragraph{Approximating the distribution of $\mu$} The downside of Monte Carlo simulation is high compute overhead; 
this is exacerbated by that \sys{} must run the simulation repeatedly online for planning, as $\bar{x}$ and $S$ can only be observed online.
Fortunately, we observe that each of $V'$ and $V''$ is close to a normal distribution with a similar cumulative distribution function (CDF). 
Hence, we approximate the CI width 
by treating $V'$ and $V''$ as normal distributions with standard deviation $\sigma(\mu)$.
Based on Equation~\ref{eq:3},
we derive $\sigma(\mu)$, the standard deviation of $\mu$, as below. 

\begin{equation}\label{eq:5}
\sigma^2(\mu) =
\begin{cases}
(\sigma^2(u_x) + \bar{x}^2)(\mu^2(e')+\sigma^2(e')) - \bar{x}^2\mu^2(e') & \quad \text{if } \bar{x} > \theta\\
\sigma^2(u_x) + \sigma^2(e'') & \quad \text{if } \bar{x} \leq \theta
\end{cases}\\
\end{equation}
\begin{flalign*}
where: \quad
\sigma^2(u_x) & = \frac{S^2}{n} \sigma(t_{n-1}) = \frac{S^2(n-1)^2}{n(n-3)^2}
\end{flalign*}
As $V'$ and $V''$ are approximated as normal distributions, we have their CI widths as:
\begin{equation}\label{eq:6}
v'_{\alpha} = z_{\alpha/2} \sigma(\mu) \;\;\;
v''_{\alpha} = z_{\alpha/2} \sigma(\mu)
\end{equation}
where $z_{\alpha/2}$ is the Z-score for the given confidence level~\cite{intro-stats}, e.g., 1.96 when $\alpha$ is 95\% and 2.576 when $\alpha$ is 99\%.

The approximation above also sheds light on how different factors affect CI widths.
For instance, higher variations ($S$) in observed object counts and larger per-frame count errors ($\sigma(e')$ or $\sigma(e'')$) contribute to $\sigma(\mu)$, resulting in wider CI widths ($v'(\alpha)$ or $v''(\alpha)$).
A larger amount of samples ($n$) reduces the CI widths.

\paragraph{From mean counts to aggregated counts}
With the above steps \sys{} estimates the mean count \textit{per frame},
 e.g., ``the average number of cars on the road at 1FPS is $[0.5 \pm 0.1]$''.
To get aggregated counts, 
\sys{} multiplies the mean by the amount of frames, 
e.g., ``the total number of cars in 30 minutes is
$ [900 \pm 180]$''. See \sect{impl} for details.

%
%
%
%

%% file: tab-notations.tex

\begin{table}[t!]
\footnotesize
\vspace*{-1em}
\begin{tabular}{@{}l@{$\;\;$}p{2.65in}@{}}
\toprule[0.7pt]
\textsf{\textrm{Notation}} & Description \\ \toprule
$x = x_1..x_n$ & 
Inexact counts on sampled frames observed by a given NN counter \\ 
$\bar{x}, S$ & The mean and standard deviation of $x$ \\ 
$\mu_x$ & A random variable representing the mean of inexact per-frame counts (from the given NN counter) on all frames \\ 
$\mu$ & A random variable representing the mean of exact per-frame counts on all frames \\ 
$E', E'', \theta$ & 
$E'$ (or $E'')$ is the distribution of the deviation between $\mu$ and $\mu_x$
when $\bar{x}$ is above (or below) a threshold $\theta$, respectively 
 \\ 
$V',V''$ & The distributions of $\mu$, when $\bar{x}$ is above (or below) $\theta$, respectively
\\ 
$v'_{\alpha},v''_{\alpha}$ & The CI widths for $\mu$ with confidence level $\alpha$, when $\bar{x}$ is above (or below) $\theta$, respectively \\ 
\midrule[0.7pt]
\end{tabular}
\caption{Notations used in \sect{design-CI}}
\label{tab:notations}
\end{table}


%% file: fig-CI-workflow.tex
\begin{figure}[t]
	\centering
	\includegraphics[width=0.4\textwidth]{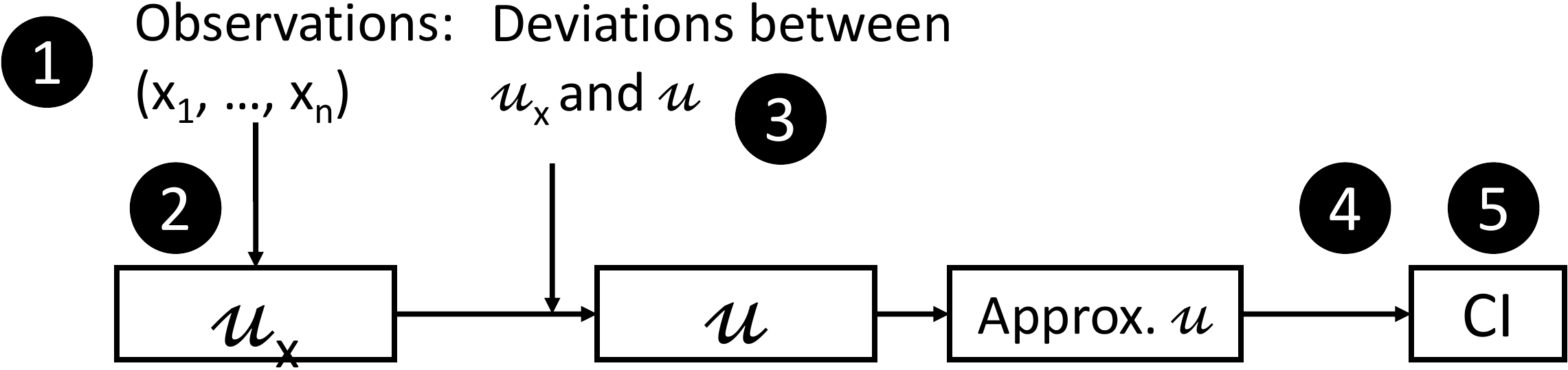}
	\caption{The workflow of deriving the CI}
	\label{fig:CI-workflow}
\end{figure}

%% file: impl.tex
\section{Implementation}\label{sec:impl}


\input{fig-hardware}

\paragraph{Heterogeneous hardware decoupling frame capture/processing}
Commodity IoT cameras are often energy-inefficient at sampling \textit{sparse} image frames: 
to capture one frame, the whole camera wakes up from deep sleep and falls back to sleep afterward, spending several seconds. 
We measured that the energy for capturing a frame is almost the same as the energy for processing the frame (YOLOv2 on Raspberry Pi 4).
While the camera may defer processing images (e.g., until window end) for amortizing the wake-up energy cost, it cannot defer periodic frame capture. 

To make periodic image capture efficient, we build a hardware prototype with a pair of heterogeneous processors, as shown in Figure~\ref{fig:hardware}. 
The prototype includes one capture unit, a microcontroller running RTOS and capturing frames periodically with rapid wakeup/suspend; and one processing unit, an application processor running Linux and waking up only to execute NN counters. 
Our evaluation \S\ref{sec:eval-hw} validates the necessity of heterogeneity.

\input{fig-video-sample}


\paragraph{NN counters}
\sys{} builds on NNPACK-accelerated darknet~\cite{nnpack-darknet} for YOLO NNs and TensorFlow~\cite{tensorflow} for other NNs.
It uses OpenCV~\cite{opencv} for image processing.

\paragraph{ROI-based instance counting}
To avoid double-counting objects in adjacent frames, a known computer vision challenge, 
our implementation adopts a common heuristics that exploit region of interest (ROI)~\cite{biswas2017automatic, tf-vehicle-det}. 
Shown in Figure~\ref{fig:video-sample}, 
an ROI for a video specifies an image region as well as $t$, the maximum time that an interesting object takes to travel through the region. 
Accordingly, the object count within a time period is the total number of objects intersecting with ROI on all the frames sampled over the time period at the intervals of $t$. 
We are aware of enhancements for mitigating double counting, e.g., by tracking objects across frames~\cite{wei2019city,shi2018geometry}. 
Such computer vision enhancements are compatible with \sys{}: 
they add per-frame compute cost that is minor compared to object detection which changes little of our core challenge: the relation between count actions and outcomes; 
they are also orthogonal to our core contributions for producing statistical results with limited energy.

%% file: fig-hardware.tex
\begin{figure}[t]
	\centering
	\includegraphics[width=0.47\textwidth]{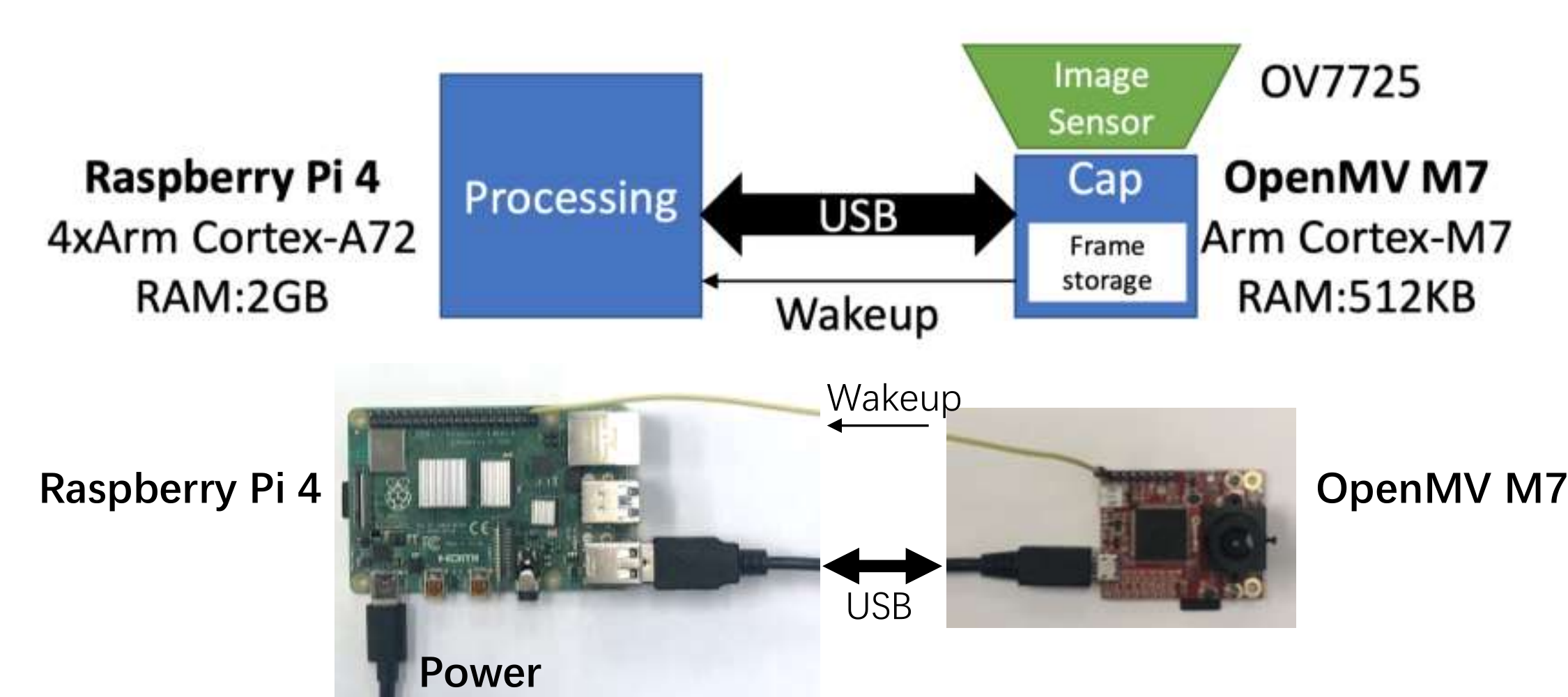}
	\caption{Our hardware prototype for testing \sys{}. The platform consists of two interconnected SoCs for frame capturing and processing, respectively.}
	\label{fig:hardware}
\end{figure}

%% file: fig-video-sample.tex
\begin{figure}[t]
	\centering
	\includegraphics[width=0.43\textwidth]{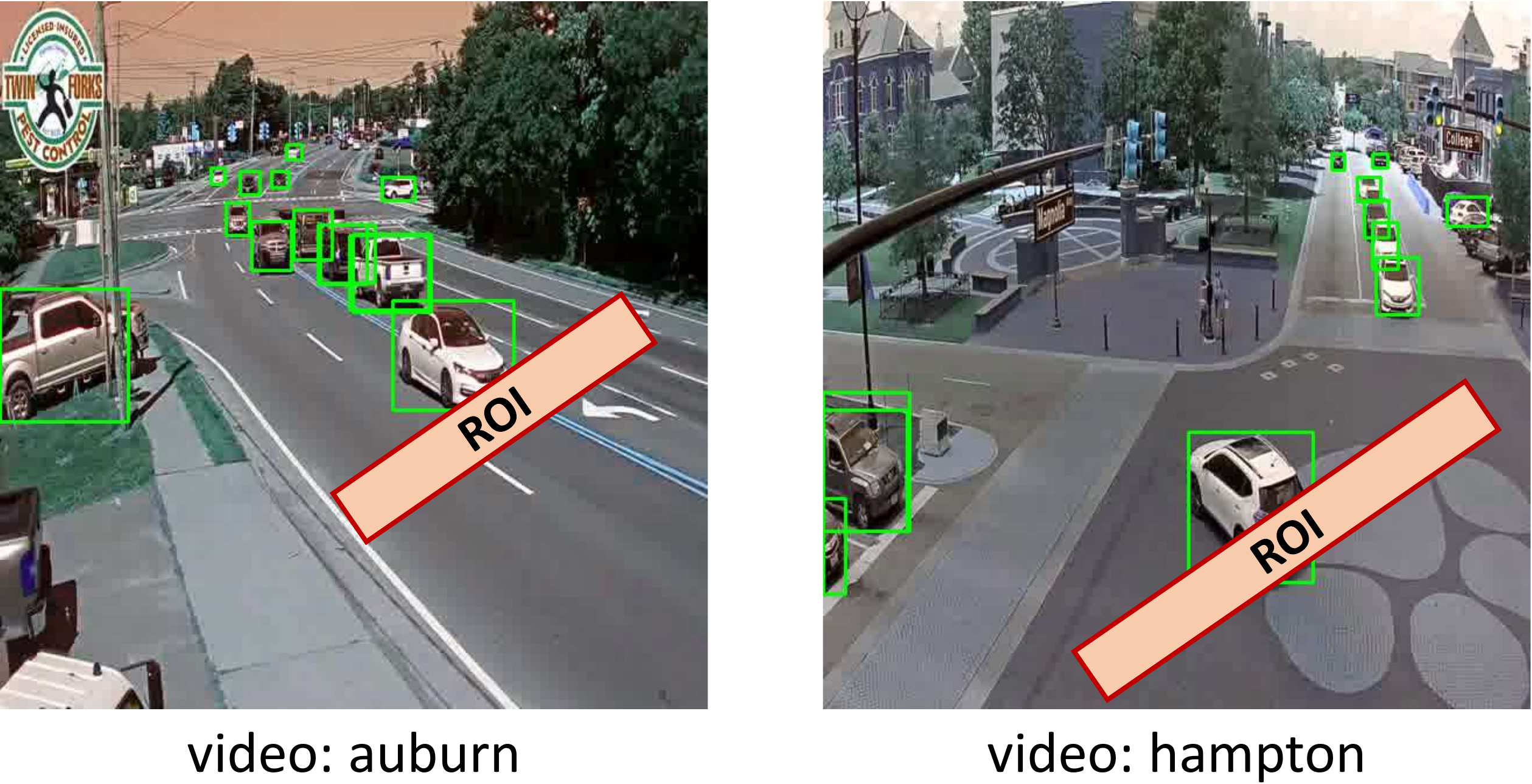}
	\caption{ROI-based object counting}
	\label{fig:video-sample}
\end{figure}

%% file: eval.tex
\section{Evaluation}\label{sec:eval}

Our evaluation answers the following questions:

\noindent $\S\ref{sec:eval-perf}$ Can \sys{} provide useful counts with valid, narrow CIs under realistic energy constraints?

\noindent $\S\ref{sec:eval-design}$ Whether the key designs of \sys{} are significant?

\noindent $\S\ref{sec:eval-hw}$ How will new hardware impact \sys{}'s performance?




\input{eval-method}

\input{eval-perf}

\input{eval-design}
\input{eval-hw}

%% file: eval-method.tex

\subsection{Methodology}\label{sec:eval-method}

\input{tab-dataset}

\input{tab-NNs}

\noindent \textbf{Videos}
We evaluate \sys{} on 5 long videos from different cameras (Table~\ref{tab:dataset}). Each video lasts 1-2 weeks; 
altogether, they constitute 1176 hours and $\thicksim$800 GB of data. 
We intentionally select the videos to cover diverse scenes, e.g., intersections and highways.
Our selection of videos is challenging to error-bounded object counting:
most frames contain few objects in ROIs; 
aggregated counts typically have small mean values and high variation.
Finding optimal count strategies for such sparse data is difficult as shown in prior work~\cite{yan2014error}.
We preprocess the videos by decoding them into 1 FPS images to accommodate our ROI-based counting ($\S$\ref{sec:impl}).
Of each video, we use the first three days of video to train our RL-based planner, and the remaining days for testing.

We report the results of counting \textit{vehicles}.
While \sys{} by design supports counting multiple object classes ($\S$\ref{sec:bkgnd-query}),
to our knowledge, there are no publicly available videos that last for days while containing diverse object classes, each class with sufficient instances for meaningful counting. 
Notably, video benchmarks popular in computer vision research only last seconds or minutes each~\cite{us-highway-dataset,wei2019city}, inadequate for exercising \sys{}. 


\revise{
Despite our best effort in finding long benchmark videos, we acknowledge the limitation in scene diversity of our video datasets. 
In the traffic videos we use, object counts are more likely to exhibit high temporal correlation, matching our rationale of using RL as discussed in \sect{design:rl}. 
Beyond traffic videos, 
we expect such temporal correlation in a variety of video scenes, e.g., cattle monitoring. 
Nevertheless, on videos where such temporal correlation is weaker, 
e.g. counting \textit{rare} wild animals, 
we expect the RL-based planner to make more misprediction. 
For such videos, while \sys{}'s characterization of count actions still holds, it would need additional heuristics for planning. 
}



\paragraph{Metrics}
To quantify \sys{}'s query answers, we report: 
\begin{myitemize}
\item \textit{CI coverage probability}: the measured chance that CIs produced by \sys{} covering the ground truth. 
The probability is expected to exceed the desired confidence level specified by users ($\S$\ref{sec:bkgnd}). 

\item \textit{Mean CI width}, normalized to the mean counts. 
For example, for a list of CIs as $\{ [ \mu_{i} \pm \delta_{i} ] \}$, the mean CI width is $(\sum{\delta_{i}}) / (\sum{\mu_i})$.
A low value indicates high overall confidence in \sys{}'s answers.

\item \textit{Mean error}, defined as $(\sum{\left|\mu_i-g_i\right|})/(\sum{g_i})$ where $g_i$ is the true count. This metric shows by how much \sys{}'s approximate counts deviate from the ground truth. 
\end{myitemize}


\paragraph{Energy} we report the whole-camera energy measured from the hardware prototype. 

\paragraph{NN counters and ground truth counts}
Table~\ref{tab:ops} lists the NN counters used in experiments and their respective energy consumptions.
Following prior work~\cite{focus,noscope,diva}, 
we treat as the ground truth the counts returned by the most expensive NN, named the ``golden'' NN counter (\code{YOLOv3}). 

\paragraph{Alternative designs} 
We compare \sys{} to the following designs:

\noindent $\bullet$ 
\OracleOp{} runs the golden NN counter with the same amount of sample frames in all windows. 


\noindent $\bullet$ 
\OptOp{} runs one \textit{single} NN counter with the \textit{same} amount of frames in all windows.
To make this design competitive, 
we set its NN counter to be the one with the best average performance over all test {\planwindow}s of a given video. 
Unlike \OracleOp{}, the counter of \OptOp{} may be different on separate videos. 
The design uses our technique ($\S$\ref{sec:design-CI}) to provide CIs.

\noindent $\bullet$
\Oracle{} uses the oracle planner descried in $\S$\ref{sec:plan:oracle}, representing the best attainable performance.
Note that \Oracle{} is built atop impractical assumptions and delays materializing aggregated counts.




\paragraph{System parameters}
We set our parameters as typically used in prior systems: 
we use 30 minutes as the \aggwindow{} length~\cite{tavana2018recent}, 24 hours as a \planwindow{}~\cite{farmbeats}, 
and 95\% as the default confidence level~\cite{blinkdb,maricq2018taming}.
As discussed in $\S$\ref{sec:bkgnd-autonomous}, the typical harvested energy from a small solar panel is 10Wh -- 30Wh, which we use in experiments.


%% file: tab-dataset.tex

\begin{table}[]
\small
\begin{tabular}{l|l|l|L{2.6cm}}
\textbf{Video}   & \textbf{Length} & \textbf{GT count} & \textbf{description} \\ 
\hline
Jackson~\cite{jackson} & 2 weeks & 1,386/3,060 & An intersection in Jackson Hole, WY \\ 
\hline
Auburn~\cite{auburn} & 1 week & 495/1,908 & Toomers Corner in Auburn, AL \\ 
\hline
Cross~\cite{cross} & 2 weeks & 329/4,412 & A three-way cross, location unknown \\ 
\hline
Taipei~\cite{taipei} & 1 week & 1,658/4,284 & An intersection in Taipei \\ 
\hline
Hampton~\cite{hampton} & 1 week & 1,267/4,824 & An interstate road in Hamptons, NY \\ \hline

\end{tabular}
\caption{Videos for evaluation.
GT count: mean/max ground-truth count of all half-hour windows.
Target object: vehicle.}
\label{tab:dataset}
\end{table}

%% file: tab-NNs.tex

\begin{table}[]
	\small
	\begin{tabular}{l|lll}
		\textbf{NN Counters}  & \textbf{Input} & \textbf{mAP} & \textbf{Energy} \\ 
		\hline
		\code{YOLOv3} (Golden, GT)~\cite{yolov3}      & 608x608    & 33.0     & 1.00         \\ 
		\code{YOLOv2}~\cite{yolov2}           & 416x416    & 21.6     & 0.22         \\ 
		\code{faster rcnn inception-v2}~\cite{faster-rcnn} & 300x300 & 28.0 & 0.40 \\ 
		\code{ssd inception-v2}~\cite{ssd-detector} & 300x300 & 24.0 & 0.08 \\ 
		\code{ssd mobilenet-v2}~\cite{mobilenetv2} & 300x300    & 22.0     & 0.05         \\ 
		\code{ssdlite mobilenet-v2}~\cite{mobilenetv2} & 300x300    & 22.0     & 0.04         \\ 
		\hline
	\end{tabular}
	\caption{The NN counters used in this work. mAP: mAP accuracy on COCO dataset~\cite{coco}.
	Energy: normalized energy as measured on RPI 4.} 
	\label{tab:ops}
\end{table}


%% file: eval-perf.tex

\subsection{End-to-End Performance}\label{sec:eval-perf}

\input{fig-eval-CI}

\input{fig-eval-end}

\noindent \textbf{\sys{} provides valid CIs}
Our experiments show that the CIs cover the ground truth count at the target coverage probability, which validates our technique for integrating errors ($\S$\ref{sec:design-CI}).
Figure~\ref{fig:eval-CICP} shows the CI coverage probability, suggesting that \sys{} has met the specified confidence level (95\%). 
The results are averaged over multiple experiment setups, i.e., energy budgets and CI width targets, as described in $\S$\ref{sec:eval-design}.

Figure~\ref{fig:eval-CICP} also shows the CI coverage by the alternative designs.
Employing our error integration technique, \Oracle{} and \OptOp{} also meet the target confidence level. 
\OracleOp{} results in noticeably lower coverage probability below the target. 
The reason is that, when running \OracleOp{} under energy constraint, the system can only afford processing a small number of frames ($<$ 30) using the golden, expensive counter. 
Such a small sample is insufficient for deriving statistically meaningful aggregates.

\noindent \textbf{\sys{} emits useful counts with limited energy.}
\sys{} produces small mean errors and mean CI widths. 
On one hand, the aggregated counts emitted by \sys{} are within 14.8\%, 12.4\%, and 11.1\% of the ground truth with an energy budget of 
10 Wh/day, 20 Wh/day, and 30 Wh/day, respectively.
On the other hand, as shown in Figure~\ref{fig:eval-end-E}, \sys{} presents mean CI widths of 22.1\%, 19.3\%, and 17.3\%, respectively.
Such results are on a par with state-of-the-art video counting approaches and analytics systems~\cite{agarwal2014knowing,liu2017video,zhang2016single,oh2019crowd}

%% file: fig-eval-CI.tex

\begin{figure}[t]
	\centering
	\includegraphics[width=0.45\textwidth]{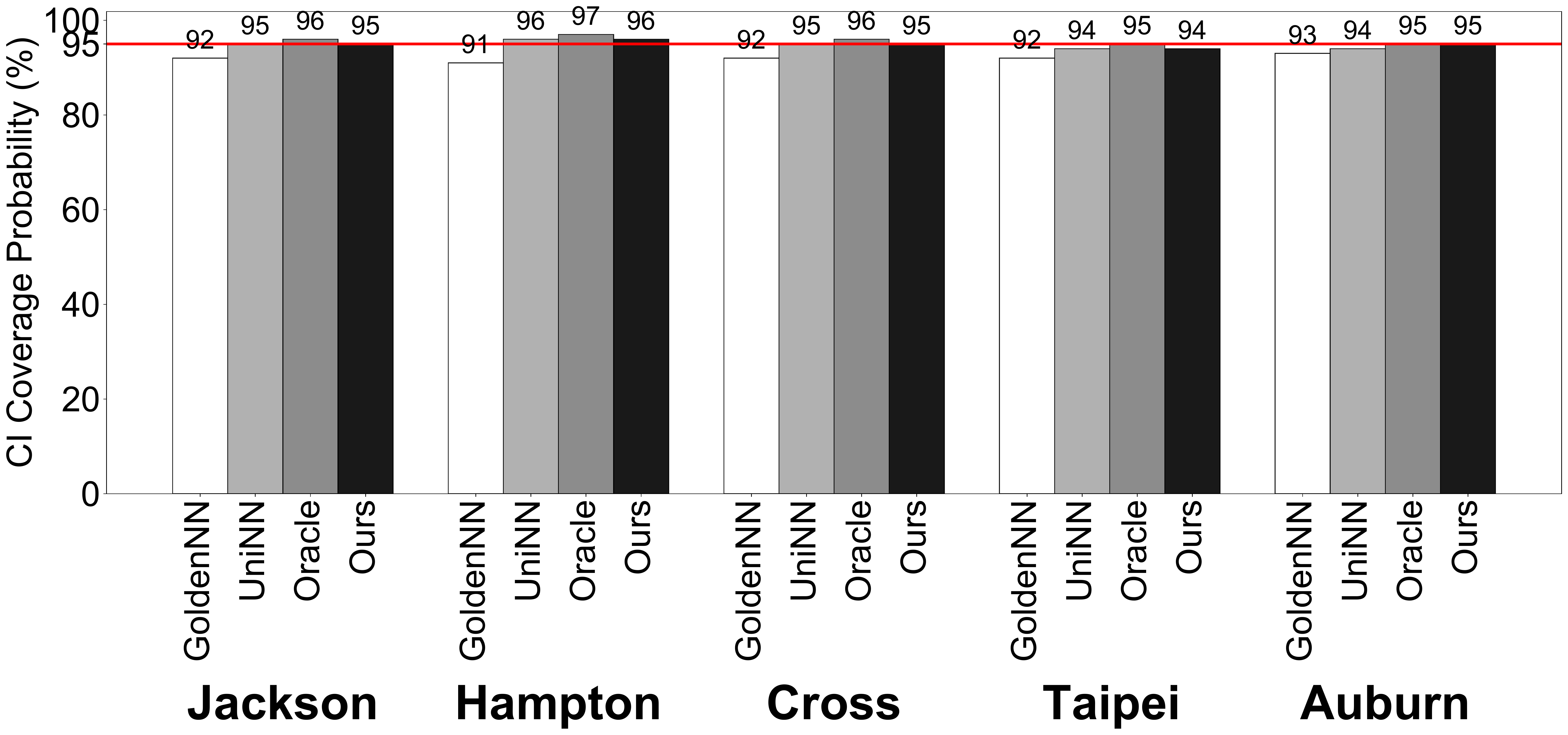}
	\caption{\sys{} produces CIs that cover the true counts at the target confidence level (95\%, the horizontal line).}
	\label{fig:eval-CICP}
\end{figure}

%% file: fig-eval-end.tex
\begin{figure*}
	\centering					
	\begin{minipage}[b]{0.18\textwidth}
		\includegraphics[width=1\textwidth]{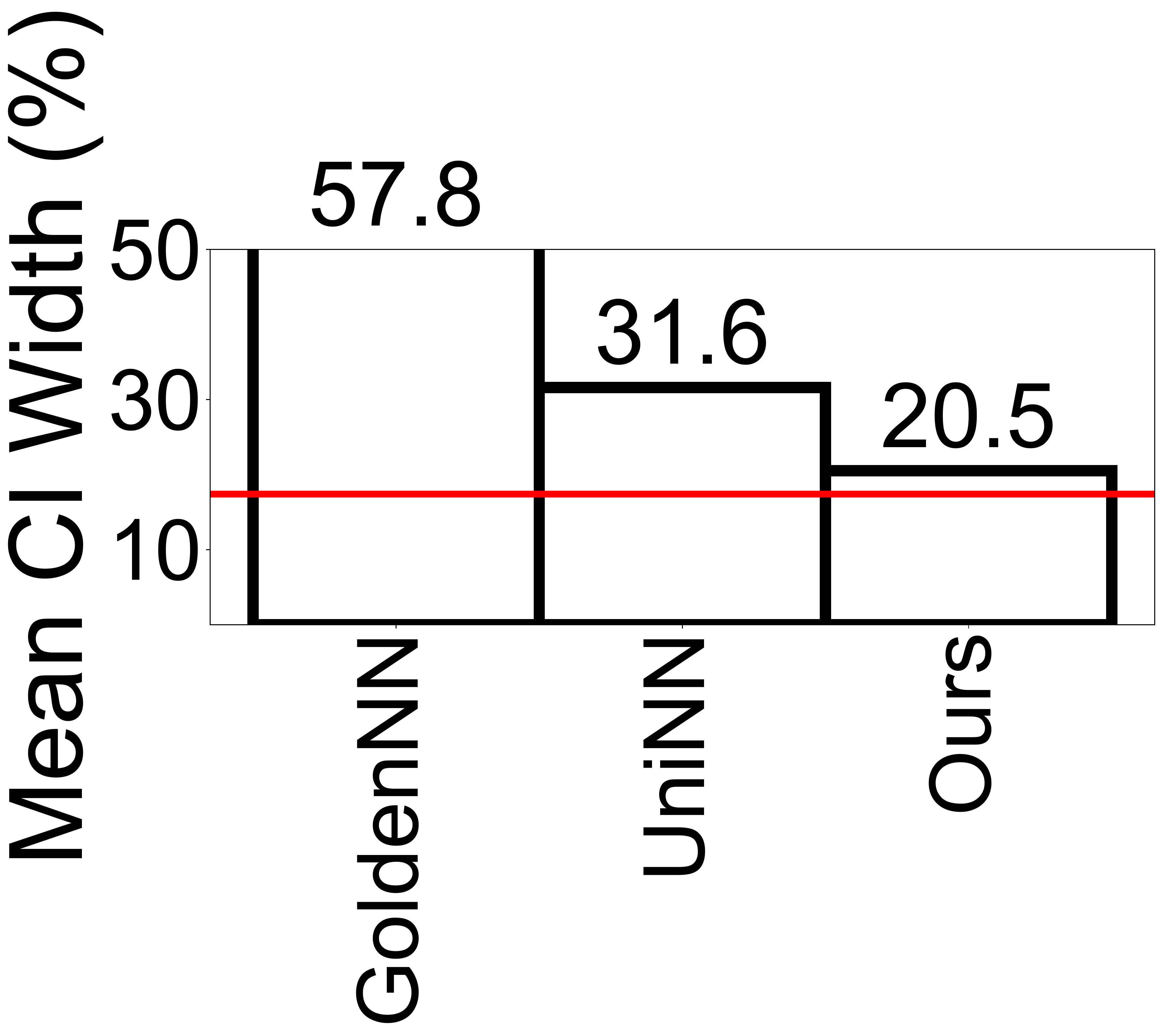}
	\end{minipage}	
	~			
	\centering				
	\begin{minipage}[b]{0.18\textwidth}
		\includegraphics[width=1\textwidth]{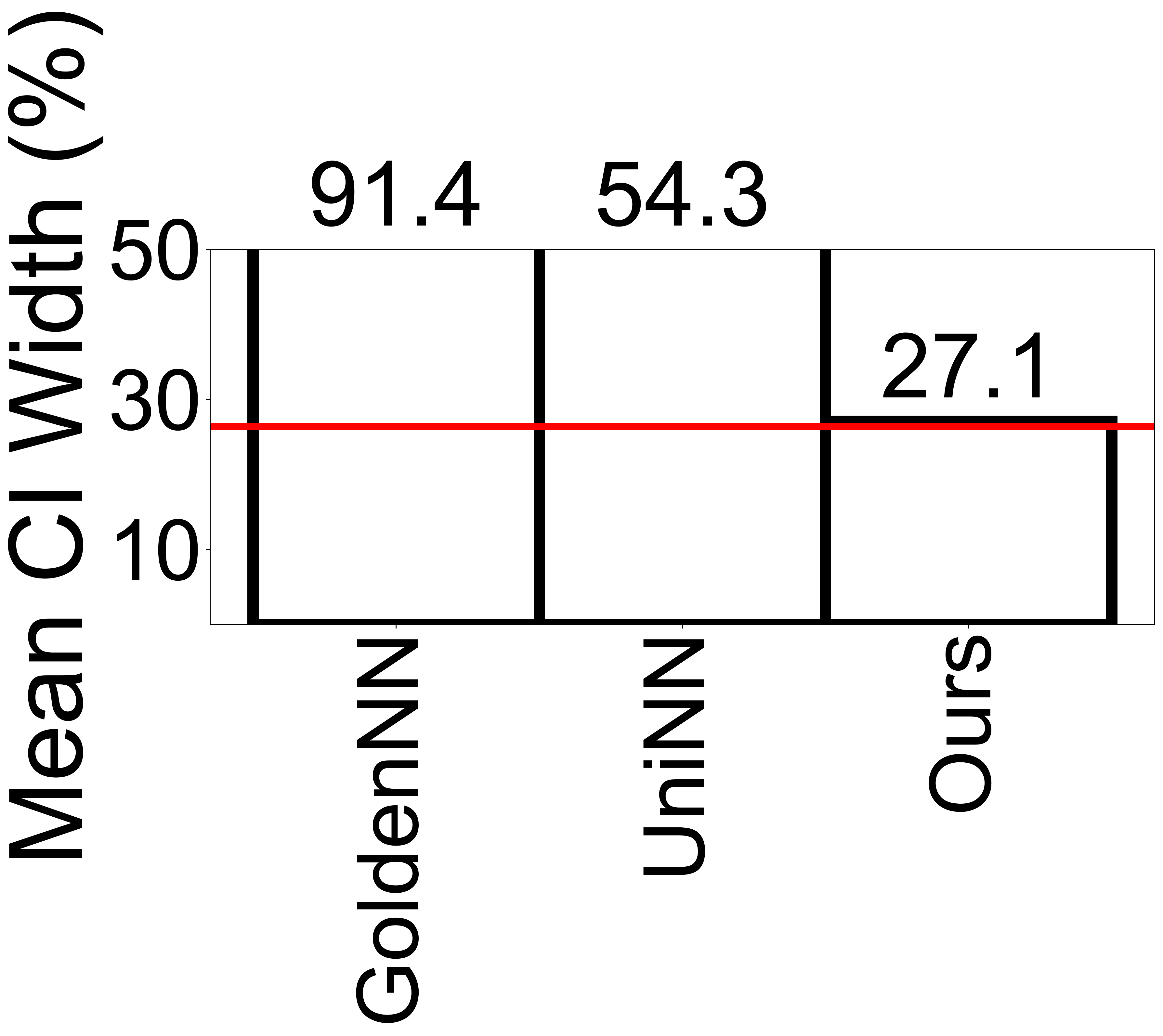}
	\end{minipage}
	~
	\centering					
	\begin{minipage}[b]{0.18\textwidth}
		\includegraphics[width=1\textwidth]{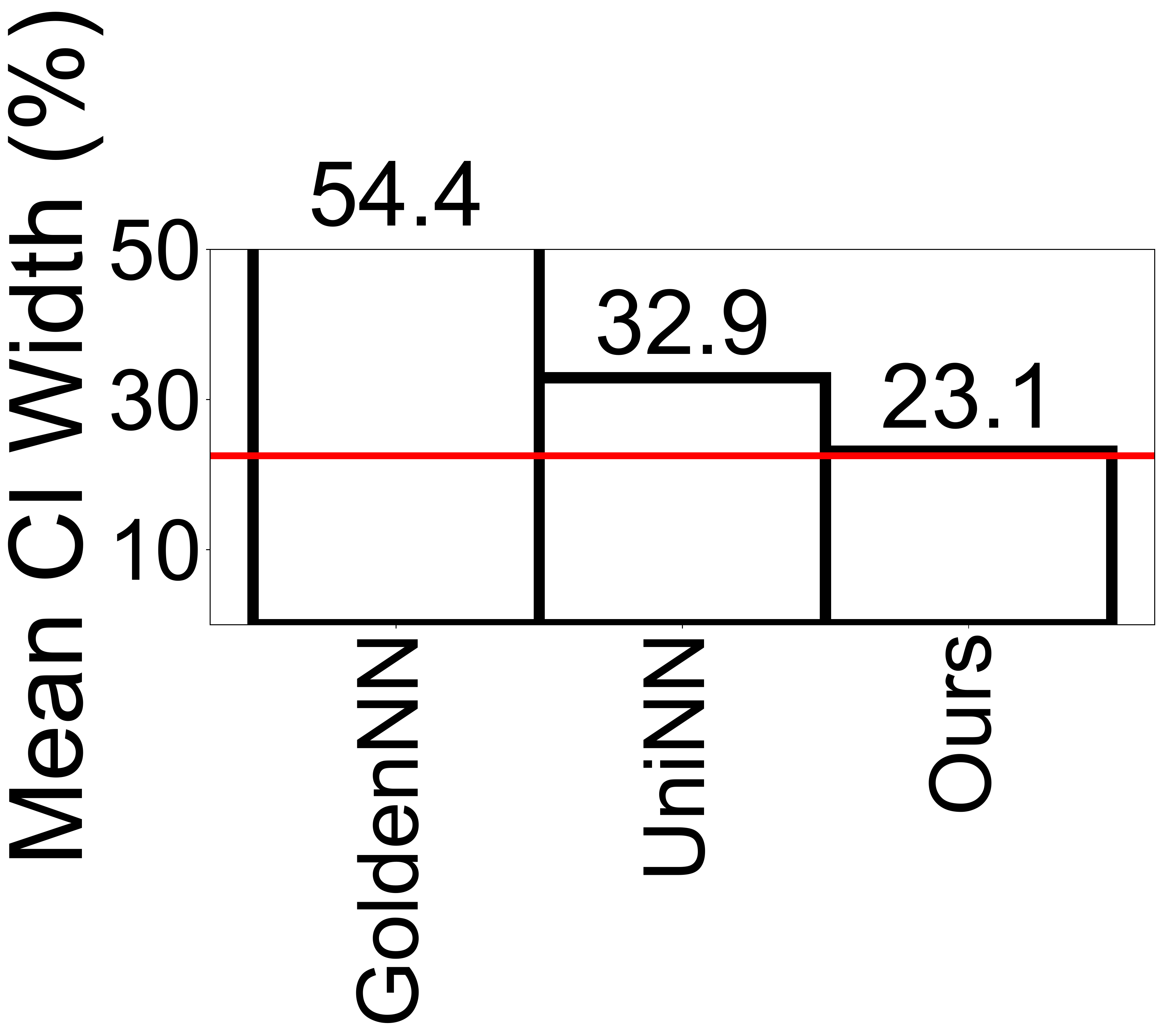}
	\end{minipage}	
	~			
	\centering				
	\begin{minipage}[b]{0.18\textwidth}
		\includegraphics[width=1\textwidth]{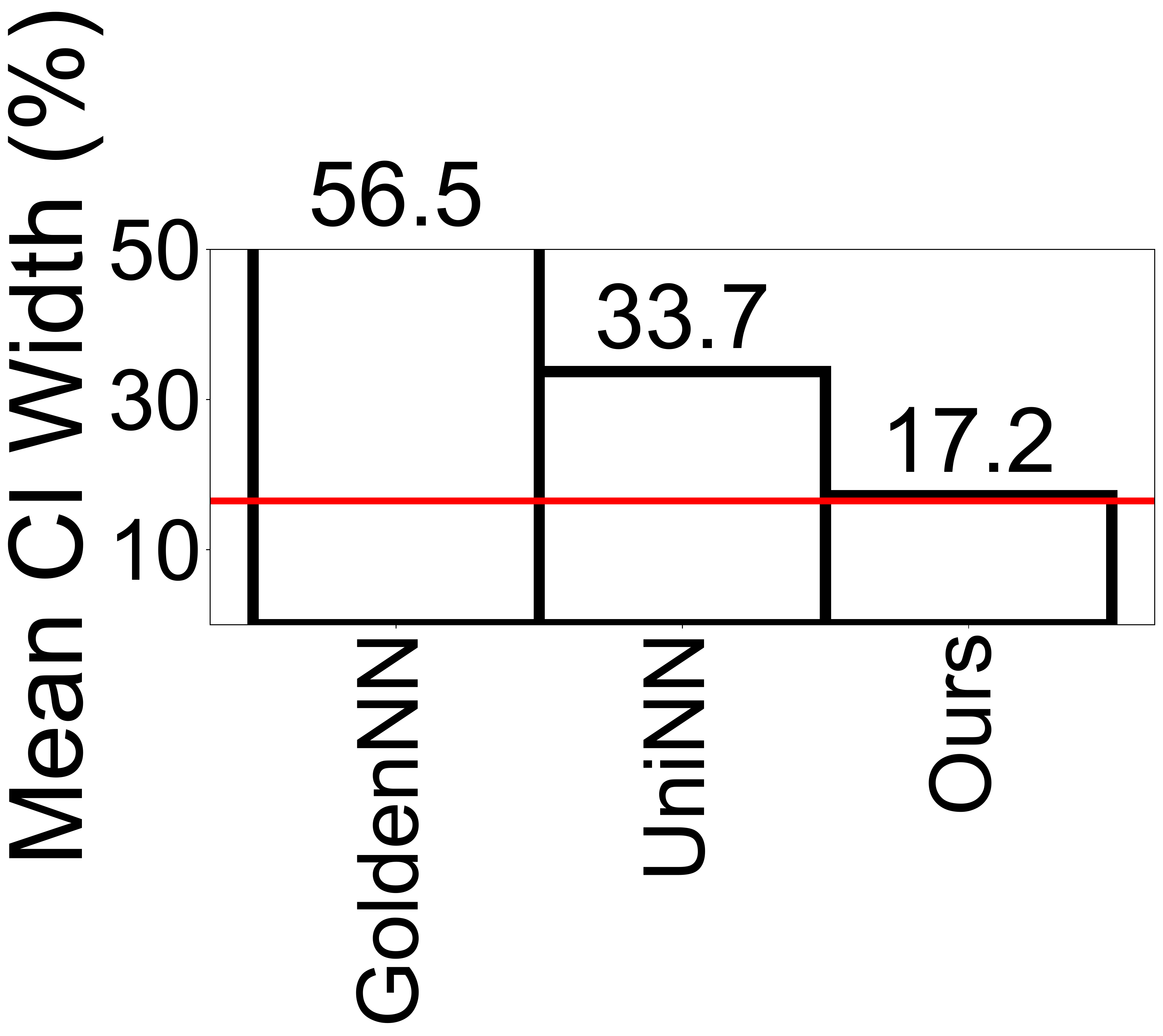}
	\end{minipage}
	~			
	\centering				
	\begin{minipage}[b]{0.18\textwidth}
		\includegraphics[width=1\textwidth]{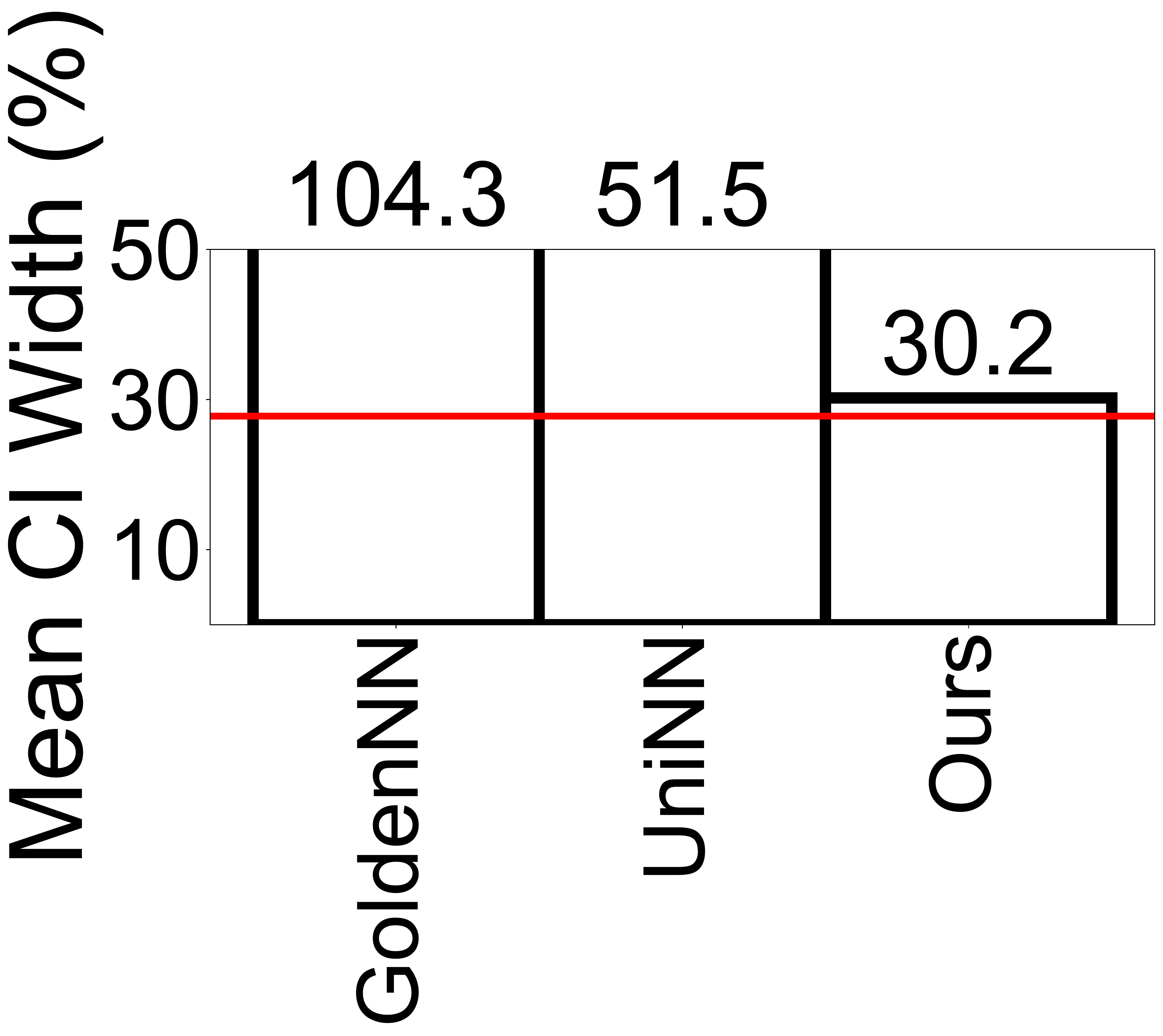}
	\end{minipage}
	
	\centering					
	\begin{minipage}[b]{0.18\textwidth}
		\includegraphics[width=1\textwidth]{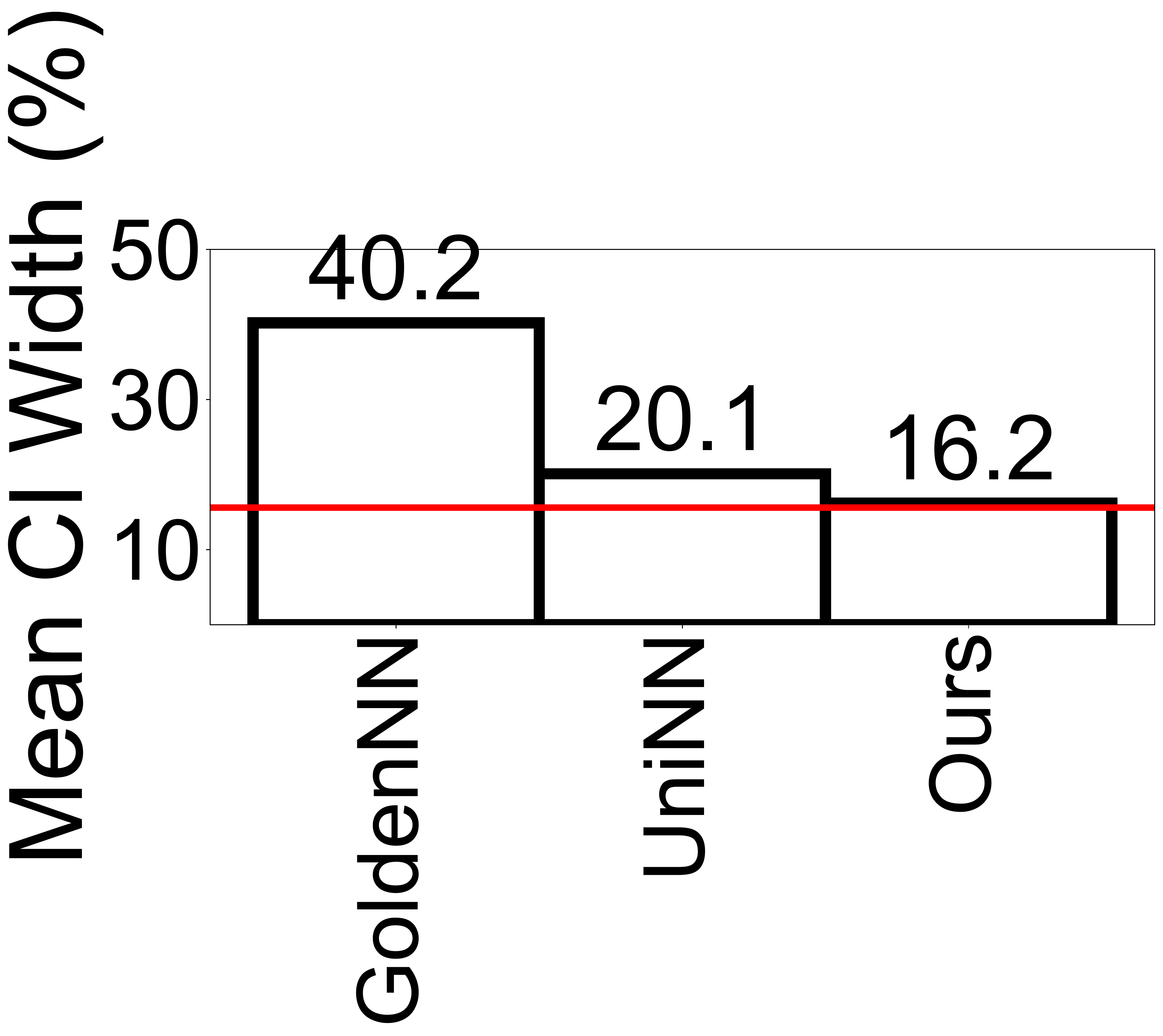}
	\end{minipage}	
	~			
	\centering				
	\begin{minipage}[b]{0.18\textwidth}
		\includegraphics[width=1\textwidth]{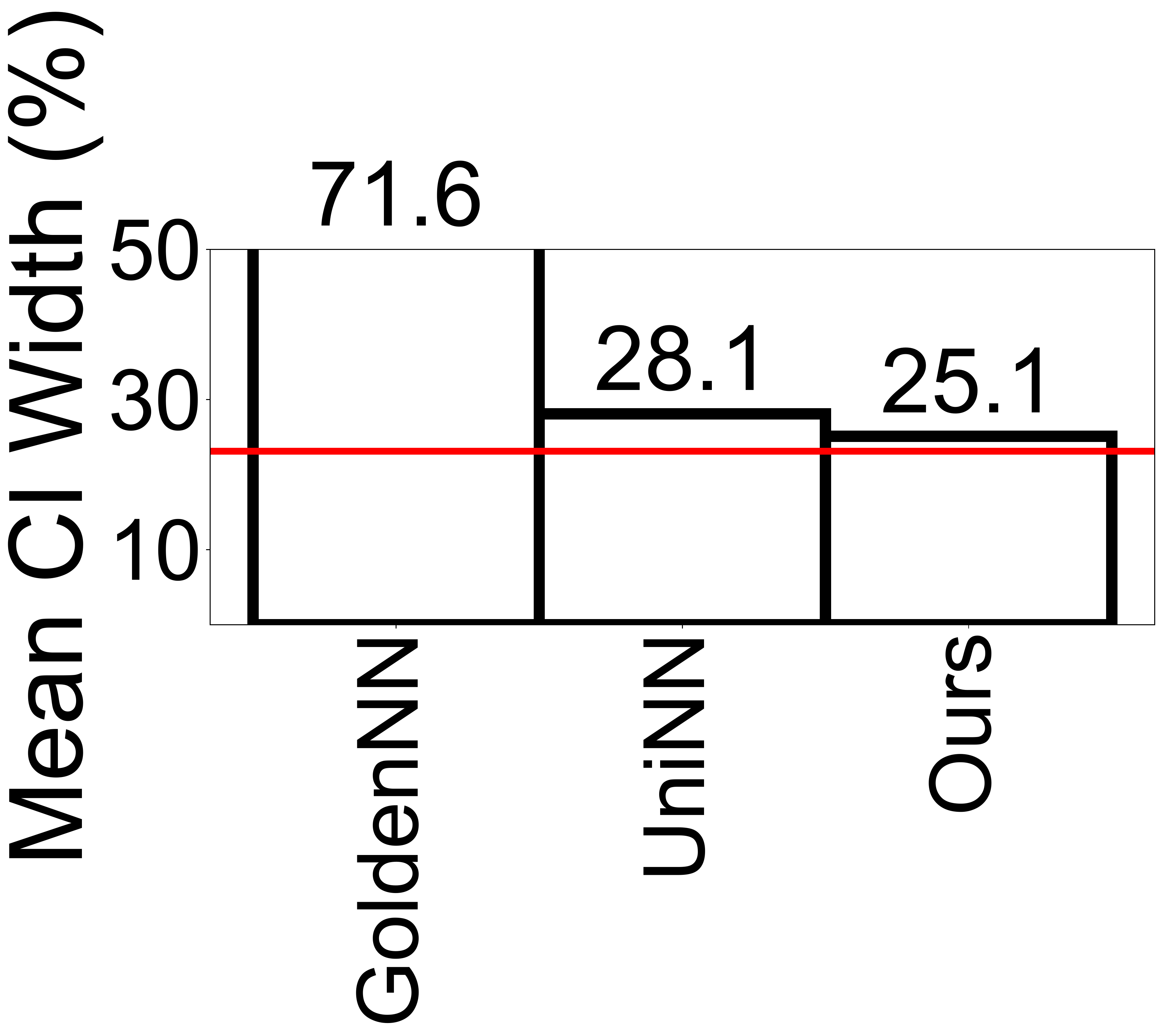}
	\end{minipage}
	~
	\centering					
	\begin{minipage}[b]{0.18\textwidth}
		\includegraphics[width=1\textwidth]{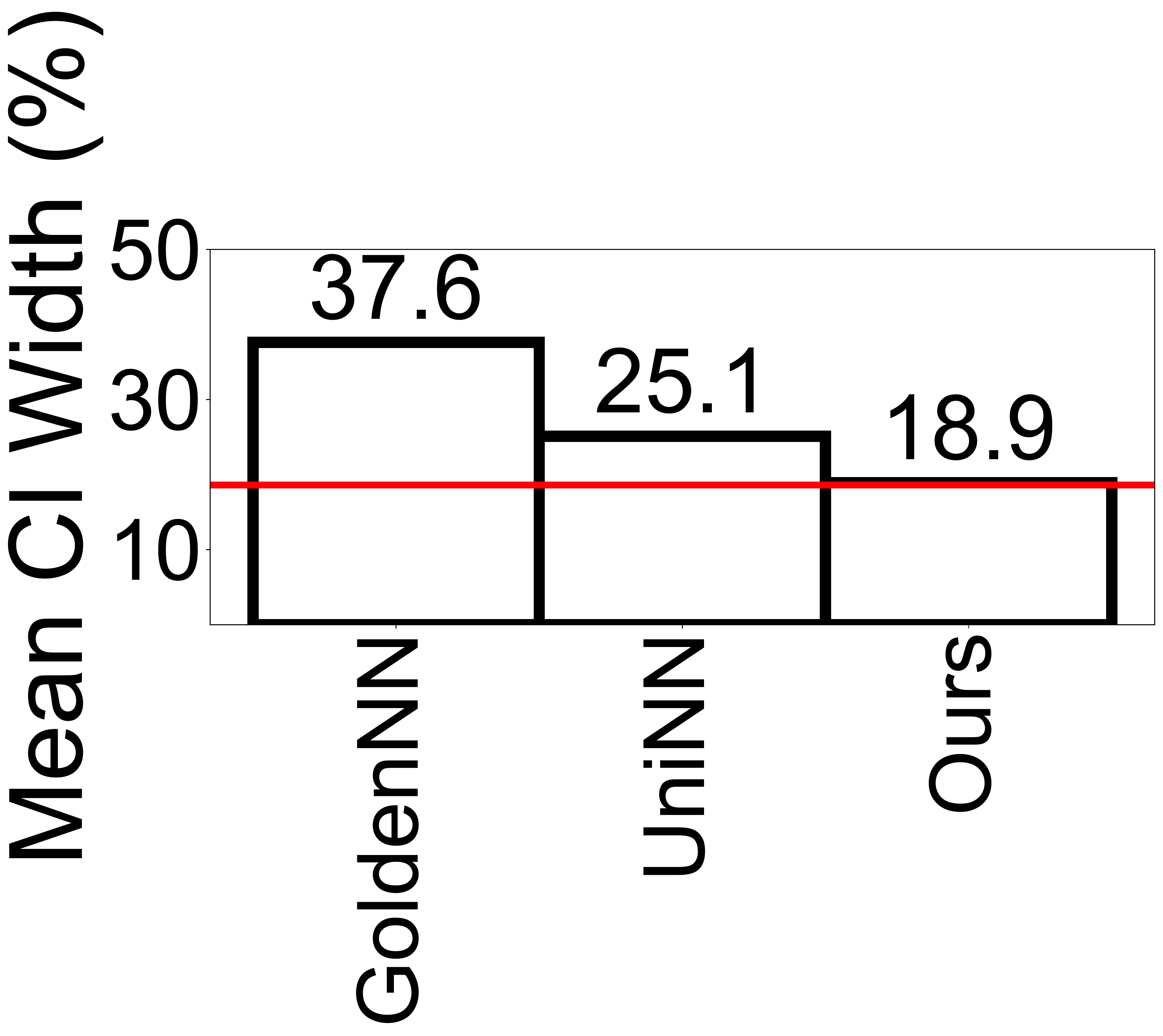}
	\end{minipage}	
	~			
	\centering				
	\begin{minipage}[b]{0.18\textwidth}
		\includegraphics[width=1\textwidth]{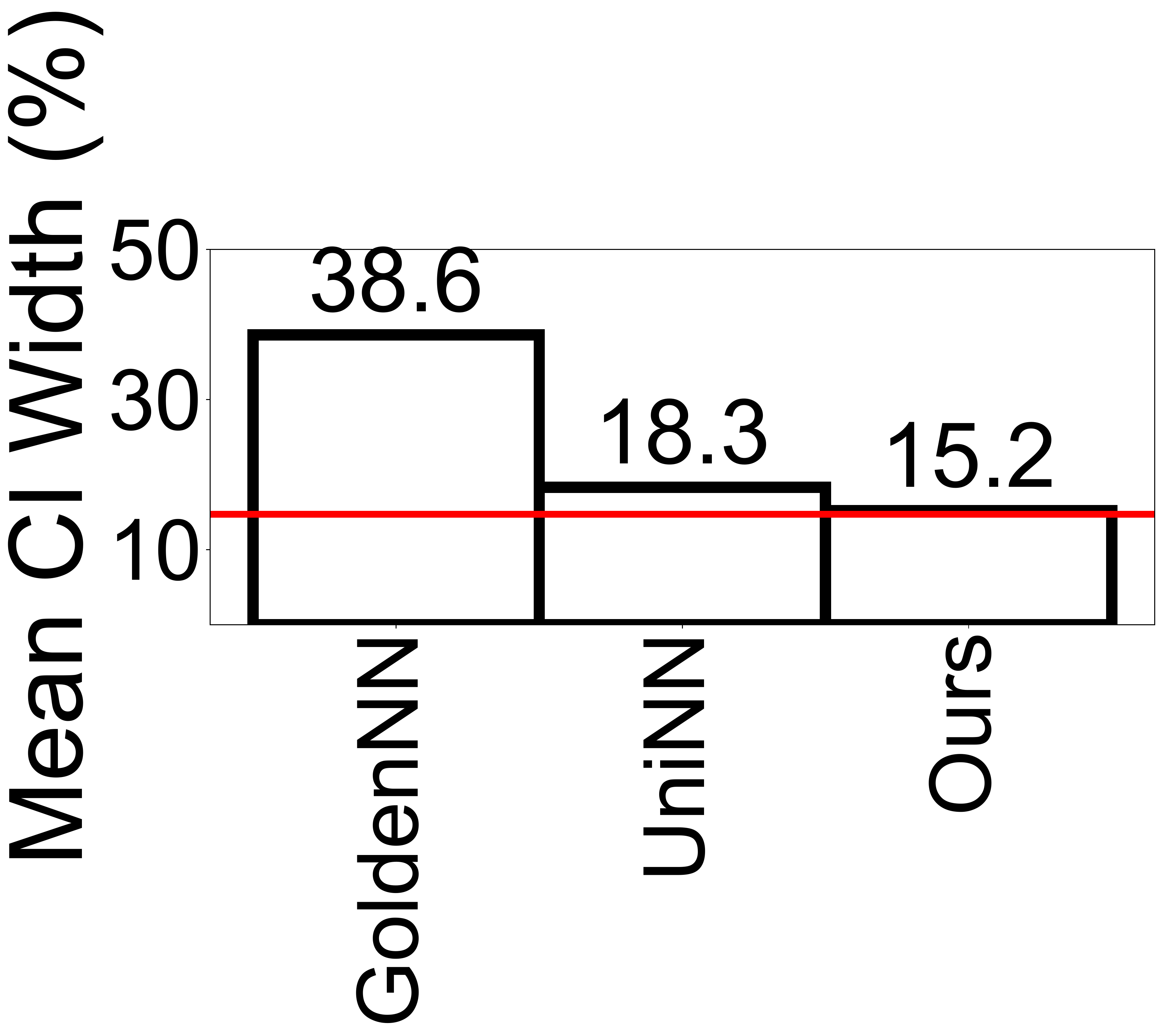}
	\end{minipage}
	~			
	\centering				
	\begin{minipage}[b]{0.18\textwidth}
		\includegraphics[width=1\textwidth]{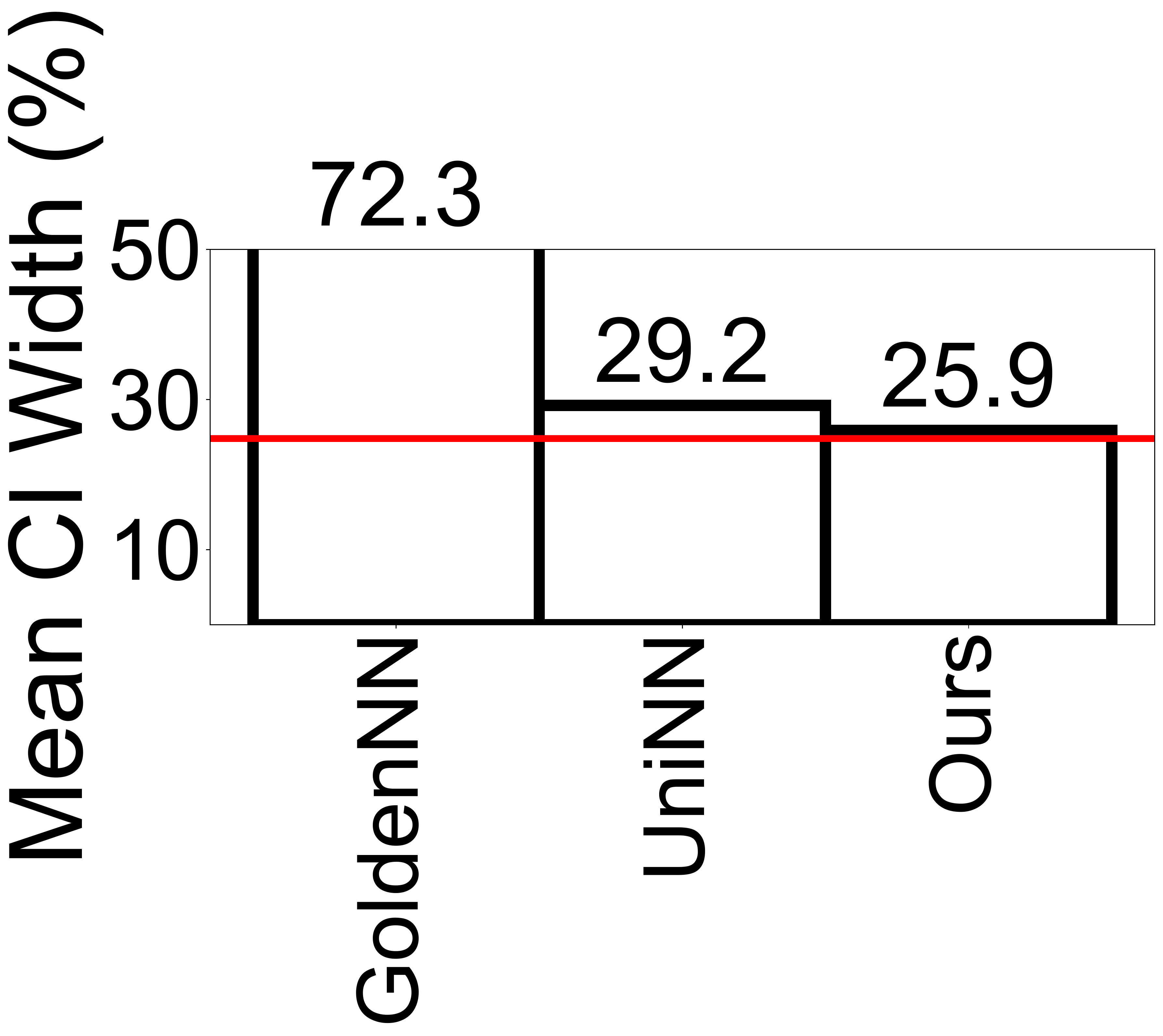}
	\end{minipage}
	
	\centering					
	\begin{minipage}[b]{0.18\textwidth}
		\includegraphics[width=1\textwidth]{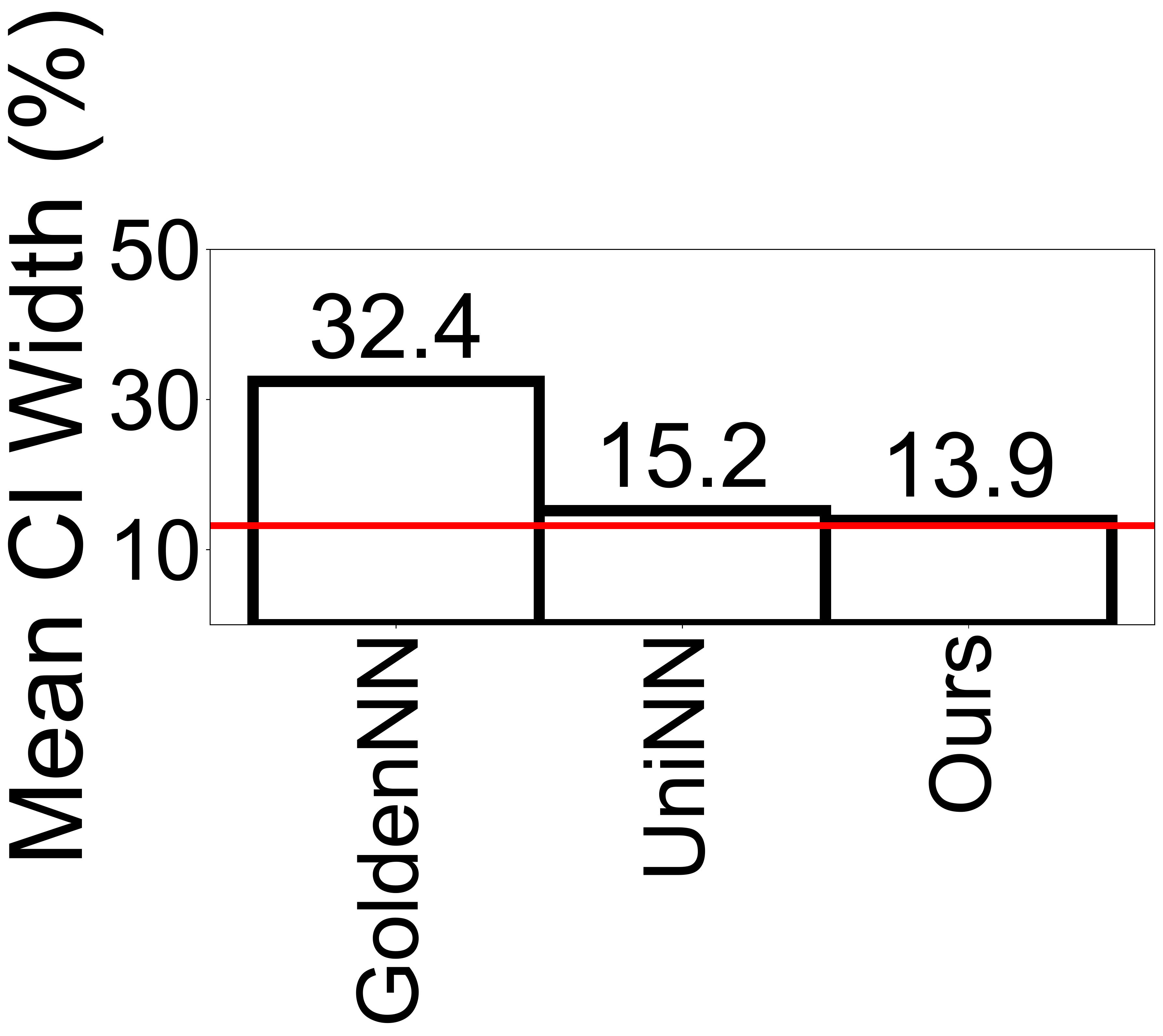}
		\subcaption{Jackson}
	\end{minipage}	
	~			
	\centering				
	\begin{minipage}[b]{0.18\textwidth}
		\includegraphics[width=1\textwidth]{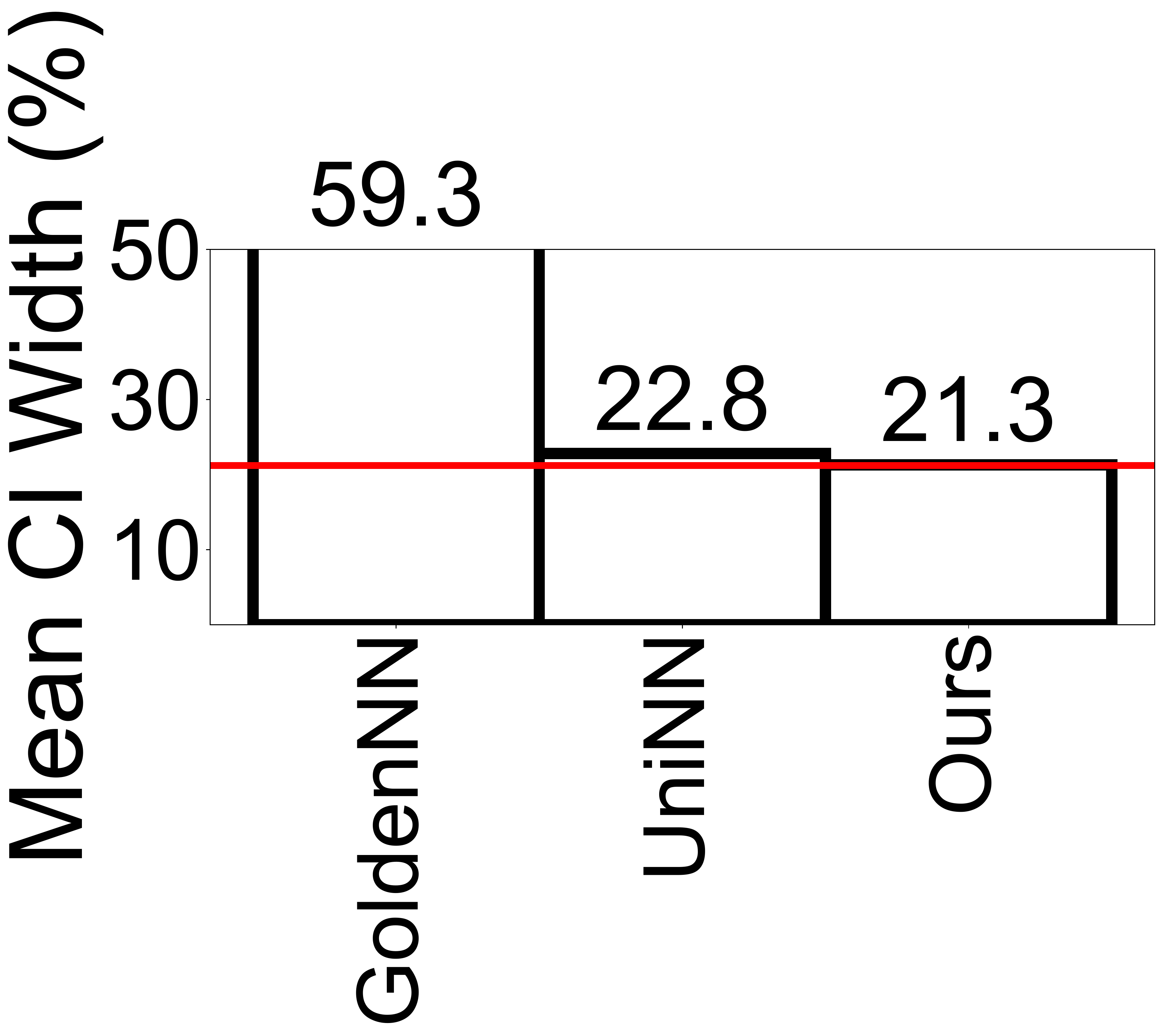}
		\subcaption{Cross}
	\end{minipage}
	~
	\centering					
	\begin{minipage}[b]{0.18\textwidth}
		\includegraphics[width=1\textwidth]{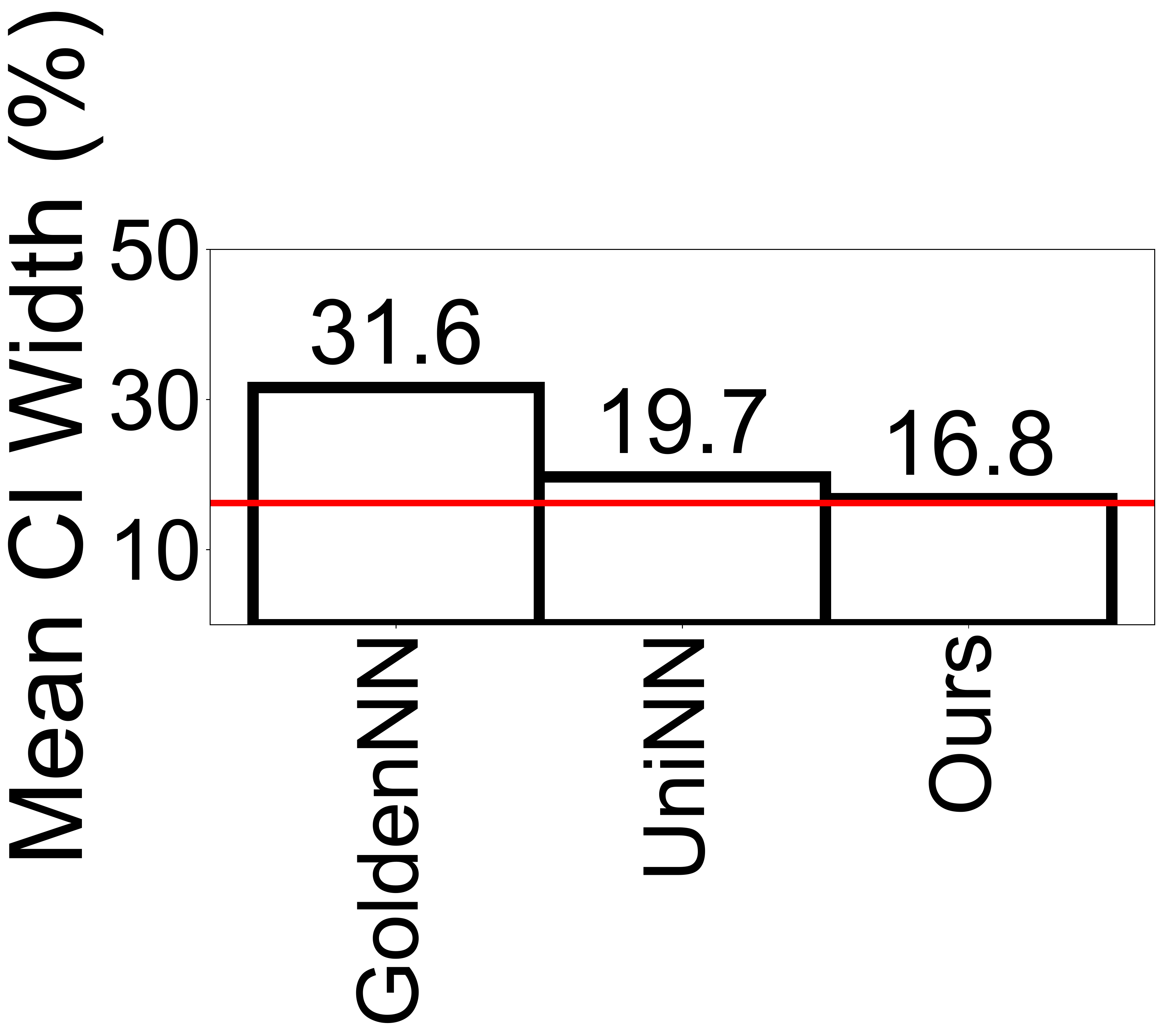}
		\subcaption{Taipei}
	\end{minipage}	
	~			
	\centering				
	\begin{minipage}[b]{0.18\textwidth}
		\includegraphics[width=1\textwidth]{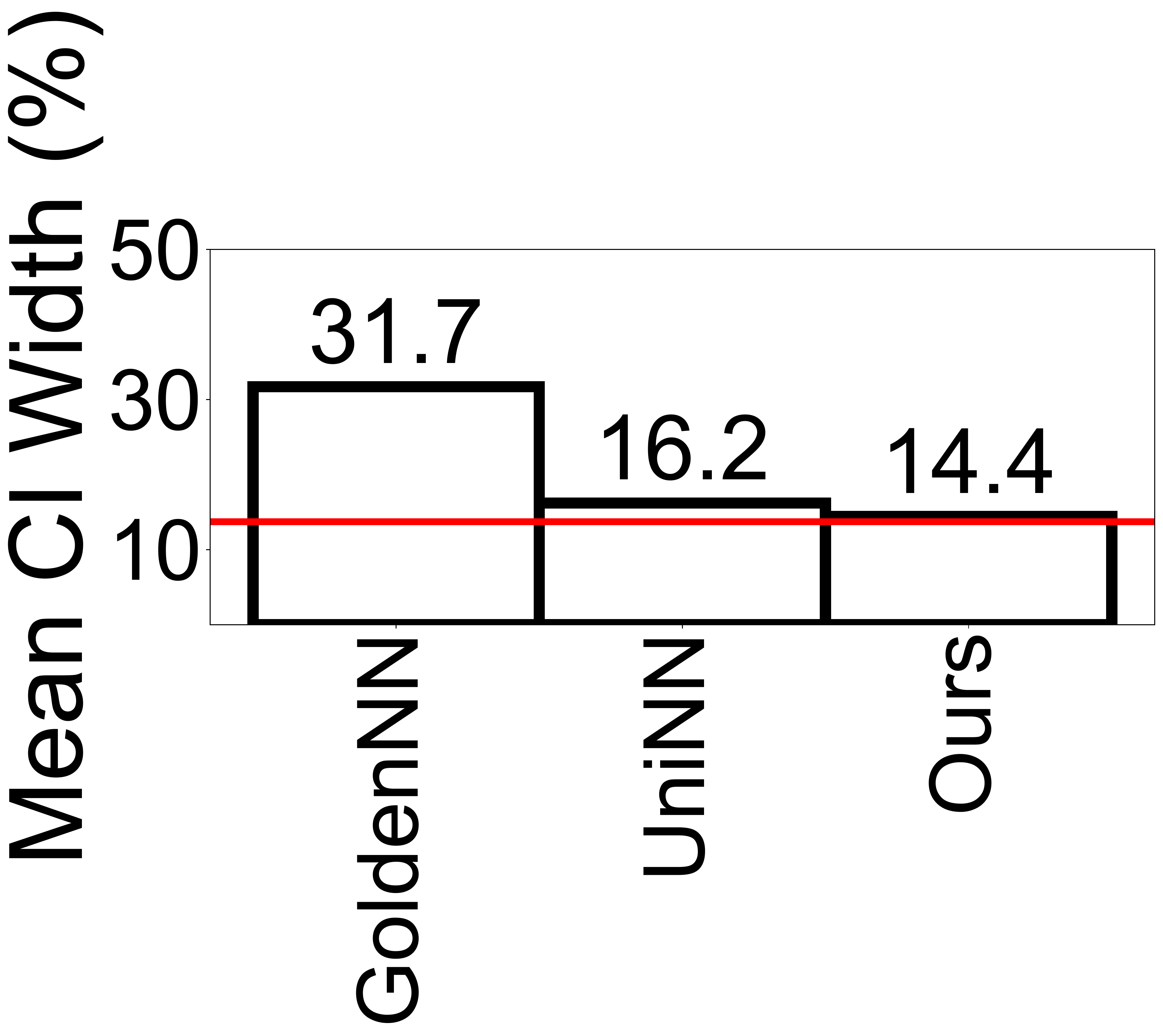}
		\subcaption{Hampton}
	\end{minipage}
	~			
	\centering				
	\begin{minipage}[b]{0.18\textwidth}
		\includegraphics[width=1\textwidth]{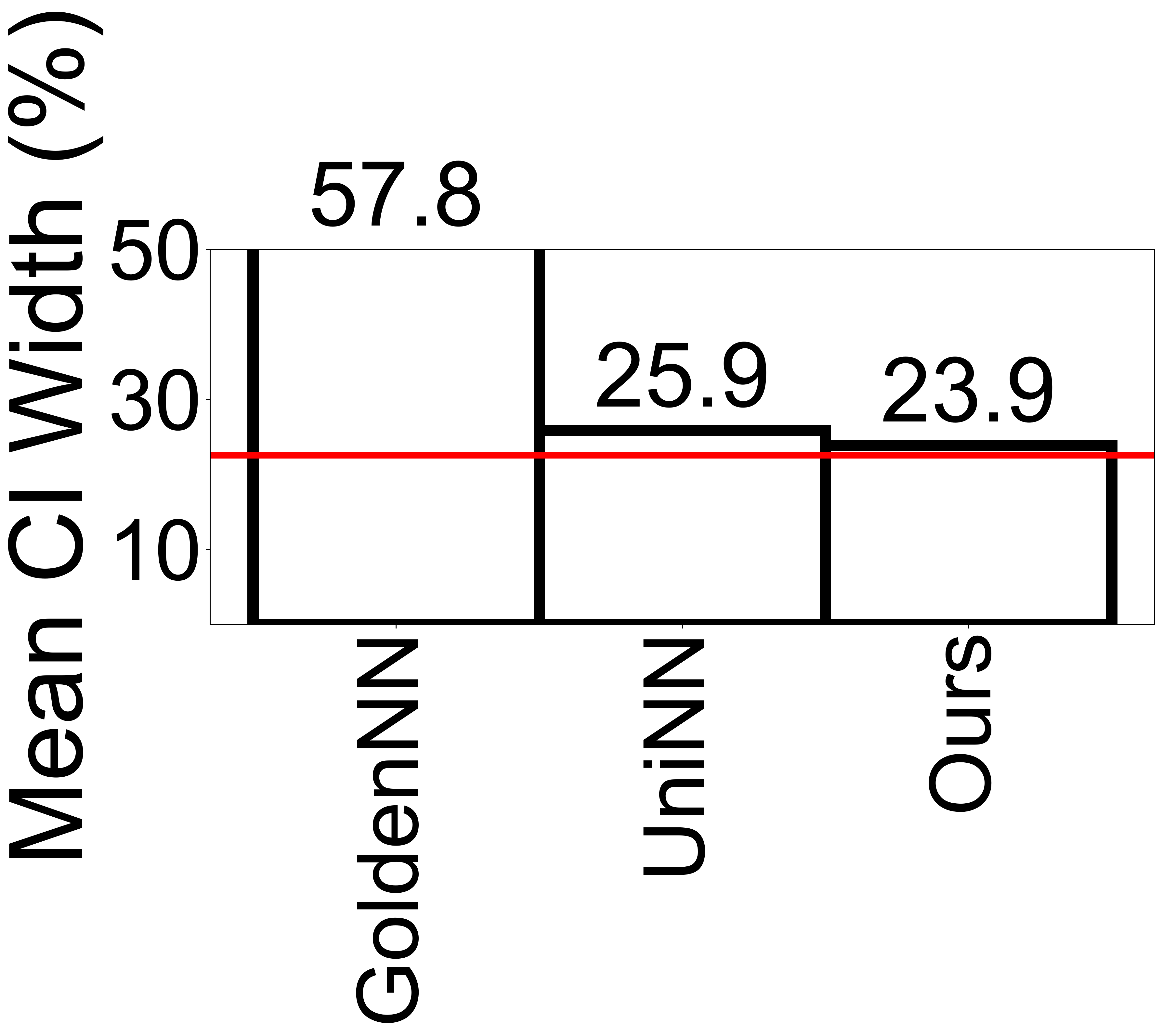}
		\subcaption{Auburn}
	\end{minipage}
	
	\caption{With given energy budgets (1st/2nd/3rd rows: 10Wh/20Wh/30Wh per day) and videos (columns), \sys{} provides narrower CIs (Y-axis) than \OracleOp{} and \OptOp, and very close to \Oracle{} (the horizontal line).}
	\label{fig:eval-end-E}
\end{figure*}

%

%% file: eval-design.tex

\subsection{Validation of Key Designs}\label{sec:eval-design}

%
%
%
%
%

\subsubsection{Exploiting diverse NN counters} 
\sys{} significantly outperforms using one counter only.
As shown in Figure~\ref{fig:eval-end-E}, compared to \OracleOp,
\sys{}'s mean CI widths are smaller by 66.6\%, 59.8\%, and 56.2\% (note: not percentage points) on average with an energy budget of 10Wh/day, 20Wh/day, and 30Wh/day, respectively.
On some videos, e.g., Auburn, \sys{}'s CI is up to 3.4$\times$ smaller.

Diving deeper, we find that
while the golden NN counter produces more exact counts per frame, the system can only sample less than 30 frames per \aggwindow. 
By contrast, \sys{} is able to pick moderately less accurate counters while sampling 5$\times$--9$\times$ more frames.
The large sample quantities outweigh the modest increase in per sample errors and result in overall higher confidence. 


\subsubsection{Heterogeneous count actions across windows}
\sys{} consistently outperforms \OptOp, a static count action optimized for all windows in a video. 
As shown in Figure~\ref{fig:eval-end-E}, compared to \OptOp,
\sys{}'s mean CI widths are smaller by 41.1\%, 16.6\%, and 9.7\% on average with an energy budget of 10Wh/day, 20Wh/day, and 30Wh/day, respectively.

The advantages of \sys{} are twofold. 
First, \OptOp{} only uses one single NN counter that performs best throughout an entire video.
By contrast, \sys{} utilizes the energy/CI front to select the proper NN counter for each individual \aggwindow.
We observe that: within one {\planwindow}, \sys{} switches among 2 -- 5 different NN counters across windows; 
across different \planwindow{}, videos, and energy budgets, 
\sys{}'s counter choices are even more diverse. 
Accordingly, \sys{} picks a wide range of sample quantities across windows, 
e.g., up to 5$\times$ difference (30--160 frames) with 10Wh/day energy budget.
Second, with a static count action \OptOp{} allocates the same energy on each window.
However, as video characteristics are disparate across time, 
the return of the same amount of energy varies substantially across different {\aggwindow}s as shown in $\S$\ref{sec:overview:problem} and Figure~\ref{fig:opt-op}.
By comparison, \sys{} identifies such disparity and adjusts energy accordingly ($\S$\ref{sec:plan}).
As we measured, the difference in energy allocations across windows can be up to 7$\times$.

\subsubsection{Imitating the oracle planner}

The confidence in \sys{}'s results is close to that of \Oracle{}, showing the efficacy of our learning approach. 
As shown in Figure~\ref{fig:eval-end-E}, compared to \Oracle,
\sys{}'s mean CI widths are only wider by 7.2\%, 4.4\%, and 4.1\% on average with an energy budget of 10/20/30 Wh/day, respectively.
Note that \Oracle{} is impractical and cannot deliver timely results as \sys{} does.

\input{fig-eval-rl}

\revise{
Zooming in, 
we find \sys{}'s planner predicts the oracle's decisions with good accuracy, as exemplified by the videos in Figure~\ref{fig:eval-rl}. 
First, the amount of per-window frame by \sys{} is within 15.5\% of the amount by the oracle. 
Second, \sys{} picks the same counters (out of six counter options) as the oracle does in 56\%--92\% of all the windows (mean/median=76\%/78\%, not shown in Figure~\ref{fig:eval-rl}).
We further examine when our planner deviates noticeably from the oracle, finding out that these are the windows i) showing \textit{irregular} variations in object counts or ii) where the oracle picks a rarely used NN counter. 
In the former situation, temporal correlation in object counts is weaker, rendering RL less effective; 
in the latter situation, as our planner does not see such rarely used NNs enough during training, it is less likely to pick them at run time. 
Note that i) even when our planner deviates from the oracle, the planner's decision is often the second optimal, incurring minor efficiency loss;
ii) the deviation in resource planning does not affect \sys{}'s statistical guarantee for analytics results. 
}

\sys{}'s planner strictly respects energy budgets allocated to it.
Recall that we tune our RL for conservative energy expenditures in order to minimize occurrences of early energy burn out before the end of a \planwindow{}, as described in \sect{design:rl}.
Thanks to such a design, \sys{} has energy leftover at the ends of 64.6\% of the  {\planwindow}s; 
on average the fraction of unused energy is as small as 9.5\%, which will be returned to the OS.
\sys{}'s per-window energy expenditures are within that of the oracle planner as low as 8.2\% on average.

%% file: fig-eval-rl.tex
\begin{figure}
	\centering					
	\begin{minipage}[b]{0.48\textwidth}
		\includegraphics[width=1\textwidth]{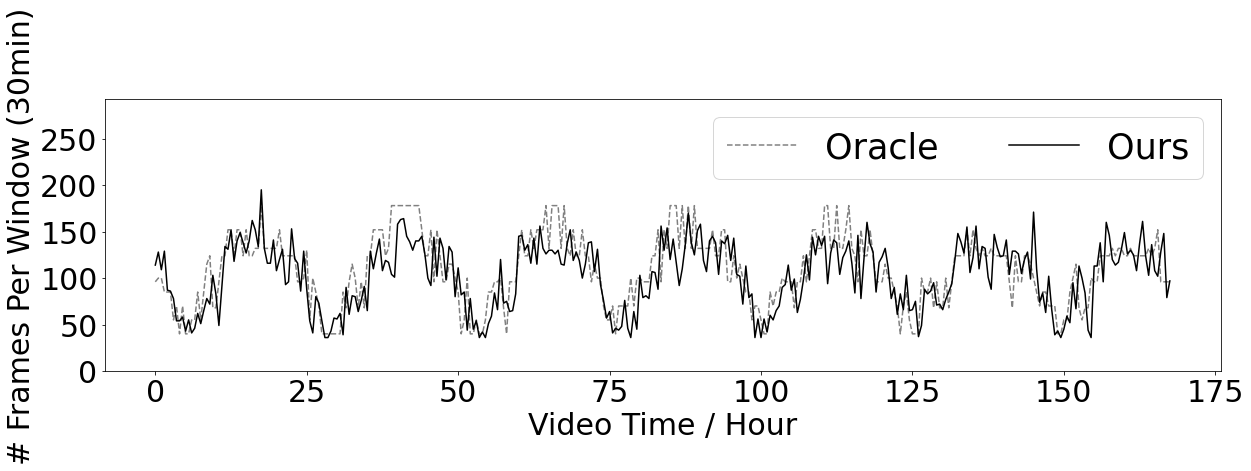}
		\subcaption{Jackson}
	\end{minipage}	
	
	\centering				
	\begin{minipage}[b]{0.48\textwidth}
		\includegraphics[width=1\textwidth]{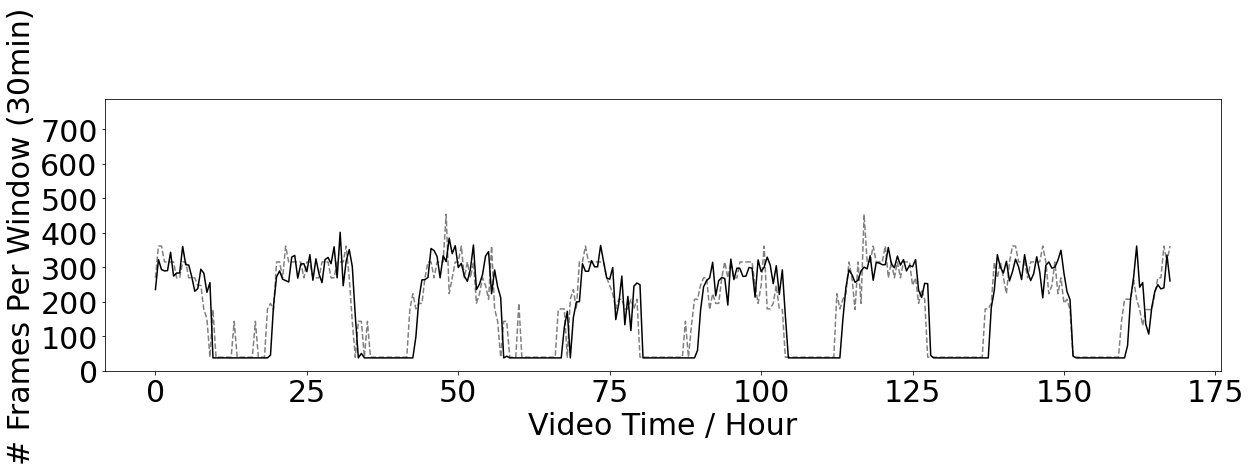}
		\subcaption{Cross}
	\end{minipage}

	\centering				
	\begin{minipage}[b]{0.48\textwidth}
		\includegraphics[width=1\textwidth]{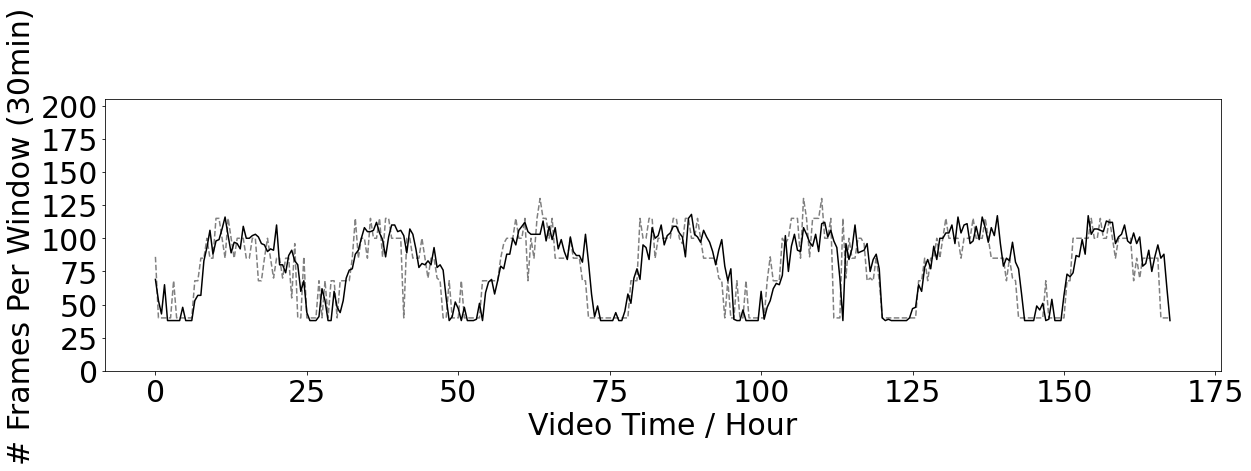}
		\subcaption{Taipei}
	\end{minipage}
	
	\caption{
	The amount of per-window frames by \sys{}'s learning-based planner as compared to the oracle planner. 	
	}
	\label{fig:eval-rl}
	
%
\end{figure}

%% file: eval-hw.tex
\subsection{The Impact of Hardware}
\label{sec:eval-hw}

\input{fig-eval-hw}

In the current \sys{} prototype, we choose commodity hardware (RPI4) that is popular for low monetary cost and programming ease. 
Next we evaluate how \sys{} behaves with different hardware.


\paragraph{Compute hardware}
\sys{} is relevant with accelerators that run NNs with higher energy efficiency; 
the extra efficiency may further expand the applicability of \sys{}. 

We test three accelerators:
Intel NCS2~\cite{intel-ncs2} (\$80), Jetson Nano~\cite{jetson-nano} (\$100), and Edge TPU~\cite{edge-tpu} (\$150).
With the results shown in Figure~\ref{fig:eval-hw}, our observations are two. 
First, compared to RPI4, more efficient accelerators reduce the mean CI width noticeably (by 22.1\%--33.1\%), primarily because \sys{} affords processing more frames per window. 
Second, even a cutting-edge accelerator (e.g., TPU) cannot reduce the CI width to near zero. 
The error mainly comes from frame sampling, showing the efficiency of the accelerator has not reached a level where \sys{} can afford processing \textit{every} frame. 
This suggests our core design -- count action planning -- to be relevant in the near future.
Third, to yield similar CI widths, the accelerators need a much smaller energy budget, hence more modest energy sources. 
For instance, with an Edge TPU, \sys{} can operate on a solar panel 16.1$\times$ smaller while producing the aggregated counts with same confidence.
Such miniaturization simplifies deployment of cameras and may suit them to low solar irradiance, or indoor environments. 

\paragraph{Image capture hardware}
Efficient image capture as in our prototype (Figure~\ref{fig:eval-hw}) matters. 
When \sys{} performs image capture with the less efficient Cortex-A72, 
we measured the per-frame energy for capture is almost 10$\times$ compared to our prototype which uses Cortex-M7 for capture. 
With less energy available for frame processing, \sys{} resorts to cheaper counters or fewer frames to sample, leading to much wider CIs (an increase of 35.6\% -- 59.0\%) as shown in Figure~\ref{fig:eval-hw} (``higher $E_{cap}$''). 

Efficient capture will be increasingly important as more efficient compute hardware emerges.
We consider ultra low-power capture hardware~\cite{cevik2015ultra} that reduces capture energy by 10$\times$
and estimate its impact as shown in Figure~\ref{fig:eval-hw} (``lower $E_{cap}$''). 
Compared to Cortex-M7 used in our prototype, the ultra low-power capture reduces the mean CI width across all compute hardware options, with more significant reduction when accelerators in use:
in the latter cases, image capture will contribute a higher fraction of system energy, or even become the bottleneck of the system energy.

%% file: fig-eval-hw.tex

\begin{figure}[t]
	\centering
	\includegraphics[width=0.42\textwidth]{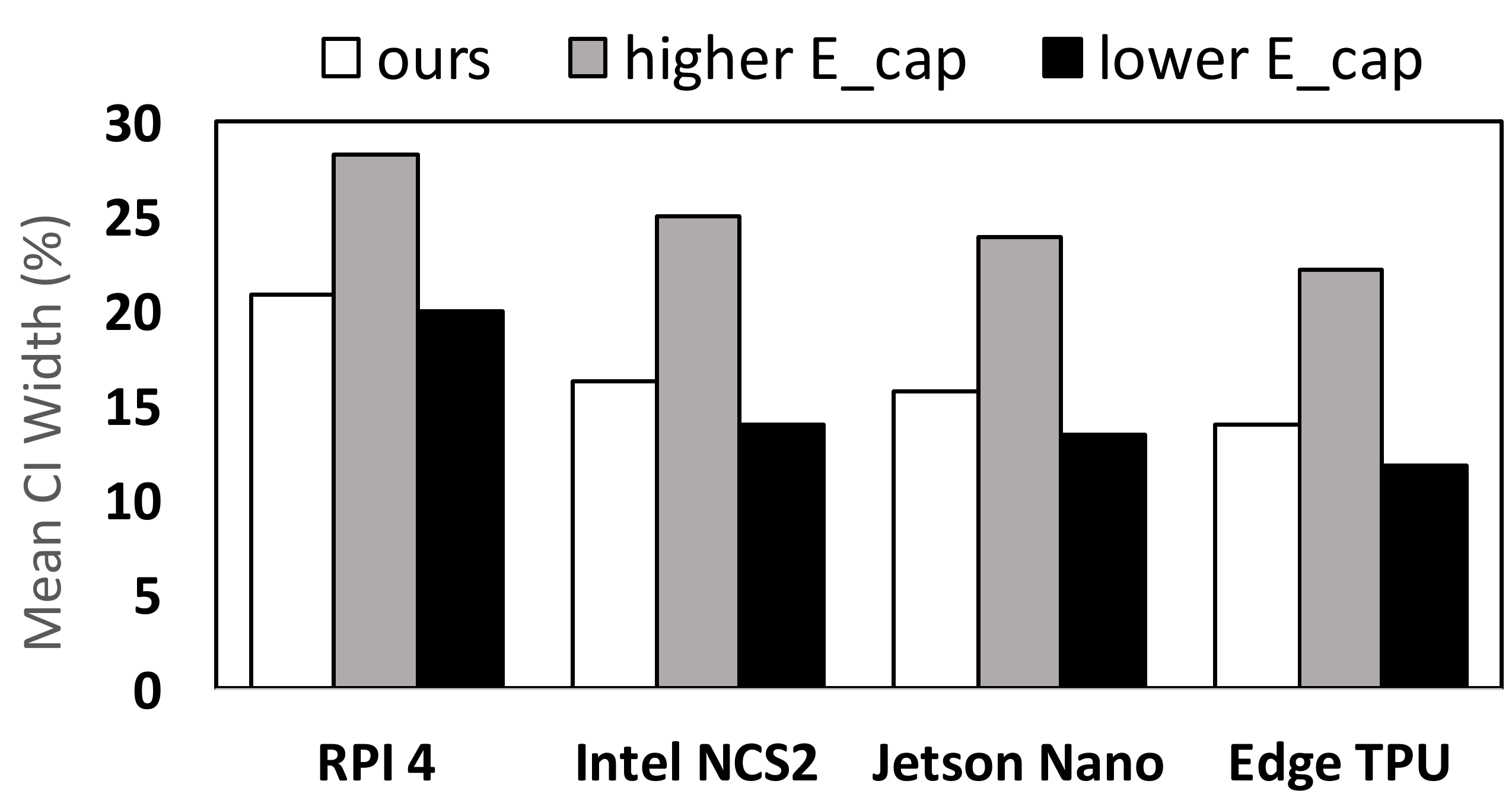}
	\caption{\sys{} on different hardware (video: Jackson, energy budget: 10Wh/day). 
		Higher $E_{cap}$ indicates using less efficient hardware to capture, while lower $E_{cap}$ indicates using an ultra low-power capture hardware.}
	\label{fig:eval-hw}
	\vspace{-1em}
\end{figure}

%% file: related.tex
\section{Related Work}\label{sec:related}

\noindent \textbf{Object counting} is a key video query. 
It has been extensively studied in the computer vision literature and shown to enable other scientific investigations ~\cite{video-public-safety,beymer1997real,naphade2017nvidia,liu2016highway,msr-video-analytics-case,yogameena2017computer,karpagavalli2013estimating,norouzzadeh2018automatically,parham2017animal,van2014nature,hodgson2016precision}.


\noindent \textbf{Energy harvesting systems}
A variety of systems, ranging from  tiny embedded devices to datacenters, operate on harvested energy 
~\cite{goiri2012greenhadoop,goiri2014designing,singh2013yank,
colin2018reconfigurable,colin2016energy,gobieski2019intelligence,farmbeats}.
\sys{} shares their motivation and may further build atop some of their mechanisms, e.g., energy budget. 
Nevertheless, these prior systems never directly address approximate visual analytics as \sys{} does.

Battery-free cameras take a radical approach toward miniaturization  ~\cite{naderiparizi2018towards,nayar2015towards,naderiparizi2015wispcam,naderiparizi2015self,solar-AI}.
With frugal energy available on device,
these cameras are often restricted to occasionally sending out captured images or running lightweight compute such as background subtraction.
By contrast, \sys{} targets battery-powered cameras (\S\ref{sec:bkgnd});
with orders of magnitude more energy, these cameras can run richer analytics built on more capable NNs. 
\sys{} therefore explores a new design space disparate from that of battery-free cameras. 

\paragraph{Specialized hardware}
Systems like 
XNOR.ai AI Camera~\cite{solar-AI} and RedEye~\cite{redeye} embrace hardware specialized for NN or vision. 
The gained efficiency may shift some design parameters of \sys{}, 
e.g., operating on smaller solar panels or smaller capacitors as discussed in Section~\ref{sec:eval-hw}. 
These systems, however, do not eliminate the need for approximate analytics, as vision algorithms are still racing to higher accuracy at higher compute expense. 


\noindent \textbf{Optimizing video analytics}
Many systems have been proposed for video analytics, 
being real-time~\cite{videostorm,chameleon,edgeeye,wang18sec,filterforward,
deepdecision,lavea,rtface,wan2019alert,rexcam} 
or retrospective~\cite{noscope,focus,scanner,vstore,diva}.
In most of these systems, compute depends on edge/cloud infrastructures, as opposed to running solely on device which is needed by autonomous cameras. 
Most, if not all, prior systems focus on per-objects results instead of statistical summaries of videos, as discussed in \sect{intro}.

Background subtraction is a common technique for skipping similar frames without full-fledged processing~\cite{noscope,focus}.
It monitors if adjacent frames captured at higher frame rate (e.g., 1FPS) contain mostly the same pixels. 
\sys{} can barely use background subtraction for skipping any frames:
\sys{} samples frames sparsely in time (e.g., one per minute);
therefore, adjacent frames often differ on substantial pixels.

\noindent \textbf{Answering query with approximation and sampling}
Approximate query processing (AQP)~\cite{garofalakis2001approximate} speeds up queries over large data.
Typical AQP approaches include online aggregation (OLA)~\cite{ola,blinkdb,olamr,pansare2011online} and offline synopses generation~\cite{acharya1999aqua,chaudhuri2007optimized,agarwal2013mergeable}.
Besides, many sampling strategies~\cite{babcock2003dynamic,yan2014error,chaudhuri2007optimized,
acharya2000congressional,sidirourgos2011sciborq} have been proposed to improve performance.
AQP systems often answer queries with approximation and bounded error as \sys{} does.
However, they are mostly designed for relational data but not videos; 
they do not run inaccurate operators (e.g., NNs) on data. 
They do not face many challenges as \sys{} does, such as integrating multiple errors and operating under energy budget. 

%% file: conclusion.tex
\section{Conclusions}\label{sec:conclusion}
\sys{} is an analytics system to answer object counting queries on autonomous cameras.
\sys{} combines inaccurate NNs and sampling technique for video queries.
It contributes a novel mechanism to make on-the-fly count decisions within and across multiple time windows. 
It takes a novel approach to integrating errors from different sources.
Tested on large videos, \sys{} provides good estimation of object counts with bounded, narrow errors.